\documentclass[12pt]{article}

\usepackage{epsfig}
\usepackage{graphicx}
\usepackage{dcolumn}
\usepackage{bm}

\date{}
\begin{document}



\newcommand{\mc}{\multicolumn}
\newcommand{\bce}{\begin{center}}
\newcommand{\ece}{\end{center}}
\newcommand{\beq}{\begin{equation}}
\newcommand{\eeq}{\end{equation}}
\newcommand{\bea}{\begin{eqnarray}}

\newcommand{\eea}{\end{eqnarray}}
\newcommand{\cont}{\nonumber\eea\bea}
\newcommand{\cl}[1]{\begin{center} {#1} \end{center}}
\newcommand{\ba}{\begin{array}}
\newcommand{\ea}{\end{array}}

\newcommand{\ab}{{\alpha\beta}}
\newcommand{\cd}{{\gamma\delta}}
\newcommand{\dc}{{\delta\gamma}}
\newcommand{\ac}{{\alpha\gamma}}
\newcommand{\bd}{{\beta\delta}}
\newcommand{\abc}{{\alpha\beta\gamma}}
\newcommand{\eps}{{\epsilon}}
\newcommand{\lam}{{\lambda}}
\newcommand{\mn}{{\mu\nu}}
\newcommand{\mpnp}{{\mu'\nu'}}
\newcommand{\Amuu}{{A_{\mu}}}
\newcommand{\Amuo}{{A^{\mu}}}
\newcommand{\Vmuu}{{V_{\mu}}}
\newcommand{\Vmuo}{{V^{\mu}}}
\newcommand{\Anuu}{{A_{\nu}}}
\newcommand{\Anuo}{{A^{\nu}}}
\newcommand{\Vnuu}{{V_{\nu}}}
\newcommand{\Vnuo}{{V^{\nu}}}
\newcommand{\Fmnu}{{F_{\mu\nu}}}
\newcommand{\Fmno}{{F^{\mu\nu}}}

\newcommand{\abcd}{{\alpha\beta\gamma\delta}}


\newcommand{\bsigma}{\mbox{\boldmath $\sigma$}}
\newcommand{\btau}{\mbox{\boldmath $\tau$}}
\newcommand{\brho}{\mbox{\boldmath $\rho$}}
\newcommand{\bpipi}{\mbox{\boldmath $\pi\pi$}}
\newcommand{\bss}{\bsigma\!\cdot\!\bsigma}
\newcommand{\btt}{\btau\!\cdot\!\btau}
\newcommand{\bnabla}{\mbox{\boldmath $\nabla$}}
\newcommand{\bphi}{\mbox{\boldmath $\tau$}}
\newcommand{\bvarphi}{\mbox{\boldmath $\rho$}}
\newcommand{\bDelta}{\mbox{\boldmath $\Delta$}}
\newcommand{\bpsi}{\mbox{\boldmath $\psi$}}
\newcommand{\bPsi}{\mbox{\boldmath $\Psi$}}
\newcommand{\bPhi}{\mbox{\boldmath $\Phi$}}
\newcommand{\bnab}{\mbox{\boldmath $\nabla$}}
\newcommand{\bpi}{\mbox{\boldmath $\pi$}}
\newcommand{\btheta}{\mbox{\boldmath $\theta$}}
\newcommand{\bkappa}{\mbox{\boldmath $\kappa$}}

\newcommand{\bA}{{\bf A}}
\newcommand{\bfe}{{\bf e}}
\newcommand{\bb}{{\bf b}}
\newcommand{\br}{{\bf r}}
\newcommand{\bj}{{\bf j}}
\newcommand{\bk}{{\bf k}}
\newcommand{\bl}{{\bf l}}
\newcommand{\bL}{{\bf L}}
\newcommand{\bM}{{\bf M}}
\newcommand{\bp}{{\bf p}}
\newcommand{\bq}{{\bf q}}
\newcommand{\bR}{{\bf R}}
\newcommand{\bs}{{\bf s}}
\newcommand{\bS}{{\bf S}}
\newcommand{\bT}{{\bf T}}
\newcommand{\bv}{{\bf v}}
\newcommand{\bV}{{\bf V}}
\newcommand{\bx}{{\bf x}}
\newcommand{\fph}{${\cal F}$}
\newcommand{\aph}{${\cal A}$}
\newcommand{\dph}{${\cal D}$}
\newcommand{\fpi}{f_\pi}
\newcommand{\mpi}{m_\pi}
\newcommand{\Tr}{{\mbox{\rm Tr}}}
\def\Qb{\overline{Q}}
\newcommand{\delu}{\partial_{\mu}}
\newcommand{\delo}{\partial^{\mu}}
%
%
\newcommand{\up}{\!\uparrow}
\newcommand{\upup}{\uparrow\uparrow}
\newcommand{\updo}{\uparrow\downarrow}
\newcommand{\uu}{$\uparrow\uparrow$}
\newcommand{\ud}{$\uparrow\downarrow$}
\newcommand{\auu}{$a^{\uparrow\uparrow}$}
\newcommand{\aud}{$a^{\uparrow\downarrow}$}
\newcommand{\pu}{p\!\uparrow}

\newcommand{\qp}{quasiparticle}
\newcommand{\sa}{scattering amplitude}
\newcommand{\ph}{particle-hole}
\newcommand{\qcd}{{\it QCD}}
\newcommand{\integ}{\int\!d}
\newcommand{\ie}{{\sl i.e.~}}
\newcommand{\etal}{{\sl et al.~}}
\newcommand{\etc}{{\sl etc.~}}
\newcommand{\rhs}{{\sl rhs~}}
\newcommand{\lhs}{{\sl lhs~}}
\newcommand{\eg}{{\sl e.g.~}}
\newcommand{\ef}{\epsilon_F}
\newcommand{\sigt}{d^2\sigma/d\Omega dE}
\newcommand{\sige}{{d^2\sigma\over d\Omega dE}}
\newcommand{\rpaeq}{\beq
\left ( \begin{array}{cc}
A&B\\
-B^*&-A^*\end{array}\right )
\left ( \begin{array}{c}
X^{(\kappa})\\Y^{(\kappa)}\end{array}\right )=E_\kappa
\left ( \begin{array}{c}
X^{(\kappa})\\Y^{(\kappa)}\end{array}\right )
\eeq}
\newcommand{\ket}[1]{| {#1} \rangle}
\newcommand{\bra}[1]{\langle {#1} |}
\newcommand{\ave}[1]{\langle {#1} \rangle}

\newcommand{\singlespace}{
    \renewcommand{\baselinestretch}{1}\large\normalsize}
\newcommand{\doublespace}{
    \renewcommand{\baselinestretch}{1.6}\large\normalsize}
\newcommand{\bftau}{\mbox{\boldmath $\tau$}}
\newcommand{\bfalpha}{\mbox{\boldmath $\alpha$}}
\newcommand{\bfgamma}{\mbox{\boldmath $\gamma$}}
\newcommand{\bfxi}{\mbox{\boldmath $\xi$}}
\newcommand{\bfbeta}{\mbox{\boldmath $\beta$}}
\newcommand{\bfeta}{\mbox{\boldmath $\eta$}}
\newcommand{\bfpi}{\mbox{\boldmath $\pi$}}
\newcommand{\bfphi}{\mbox{\boldmath $\phi$}}
\newcommand{\bfR}{\mbox{\boldmath ${\cal R}$}}
\newcommand{\bfL}{\mbox{\boldmath ${\cal L}$}}
\newcommand{\bfM}{\mbox{\boldmath ${\cal M}$}}
\def\dblint{\mathop{\rlap{\hbox{$\displaystyle\!\int\!\!\!\!\!\int$}}
    \hbox{$\bigcirc$}}}
\def\ut#1{$\underline{\smash{\vphantom{y}\hbox{#1}}}$}

\def\UNITY{{\bf 1\! |}}
\def\Pom{{\bf I\!P}}
\def\lsim{\mathrel{\rlap{\lower4pt\hbox{\hskip1pt$\sim$}}
    \raise1pt\hbox{$<$}}}         
\def\gsim{\mathrel{\rlap{\lower4pt\hbox{\hskip1pt$\sim$}}
    \raise1pt\hbox{$>$}}}         
\def\beq{\begin{equation}}
\def\eeq{\end{equation}}
\def\bea{\begin{eqnarray}}
\def\eea{\end{eqnarray}}


\begin{center}
{\Large\bf Nonlinear $k_{\perp}$-factorization for Forward Dijets in DIS off Nuclei
in the Saturation regime}\\ \vspace{1cm}
 { \bf N.N. Nikolaev$^{a,b)}$,
W. Sch\"afer$^{a)}$, B.G. Zakharov$^{b)}$, V.R.
Zoller$^{c)}$\medskip\\  }

$^{a)}$ Institut f. Kernphysik, Forschungszentrum J\"ulich, D-52425 J\"ulich, Germany\\
$^{b)}$ L.D.Landau Institute for Theoretical Physics, Chernogolovka, Russia\\
$^{c)}$ Institute for Theoretical and Experimental Physics, Moscow, Russia\vspace{1cm} \\

{\bf Abstract\\    }

\end{center}
{\small We develop the QCD description of the breakup of
photons into forward dijets in small-$x$ deep inelastic scattering off
nuclei in the saturation regime. Based on the
color dipole approach, we derive a multiple scattering expansion
for intranuclear distortions of the jet-jet transverse momentum
spectrum. A special
attention is paid to the non-Abelian aspects of the propagation
of color dipoles in a nuclear medium. We report a nonlinear
$k_{\perp}$-factorization formula for the breakup of photons into
dijets in terms of the collective Weizs\"acker-Williams (WW) glue
of nuclei as defined in ref. \cite{Saturation,NSSdijet}. For hard 
dijets with the transverse momenta above the saturation
scale the azimuthal decorrelation  (acoplanarity)
momentum is of the order of the nuclear saturation momentum $Q_A$.
For minijets with the transverse momentum below the saturation
scale the nonlinear $k_{\perp}$-factorization predicts a
complete disappearance of the jet-jet correlation. We comment on a possible
relevance of the nuclear
decorrelation of jets to the experimental data
from the STAR-RHIC Collaboration. \pagebreak\\}



\section{Introduction}

From the parton model point of view the opacity of heavy nuclei
to high energy projectiles entails a highly nonlinear relationship
between the parton densities of free nucleons and nuclei. The
trademark of the conventional pQCD factorization theorems for
hard interactions of leptons and hadrons is that the hard
scattering observables
are linear
functionals of the appropriate parton densities in the projectile
and target \cite{Textbook}.
The parton model interpretation of hard phenomena in
ultrarelativistic heavy ion collisions calls upon the 
understanding of factorization properties 
in the nonlinear regime. A priori it
is not obvious that one can define nuclear parton densities such
that they enter different observables in a universal manner.
Indeed, opacity of nuclei brings in a new large scale $Q_A$
which separates the regimes of opaque nuclei and weak attenuation
\cite{Mueller,Mueller1,McLerran,Saturation}.
Furthermore, for  parton momenta below the saturation scale
$Q_A$
the evolution of sea from gluons was shown to be dominated
by the anti-collinear, anti-DGLAP splitting \cite{Saturation}.  
  In our early studies \cite{Saturation,NSSdijet} we  have demonstrated
that such observables as the amplitude of coherent hard
diffractive breakup
of a projectile on a nucleus or the transverse momentum distribution
of forward quark and antiquark jets in deep inelastic
scattering (DIS) off nucleus
and/or the sea parton density of
nuclei can be cast in precisely the same $k_{\perp}$-factorization
form as for a free nucleon target. Specifically, one only
needs to substitute the unintegrated
gluon structure function (SF) of the free nucleon by the
collective nuclear Weizs\"acker-Williams (WW)
unintegrated nuclear glue, which
is an expansion over the collective
gluon SF of spatially overlapping nucleons of the
Lorentz-contracted ultrarelativistic 
nucleus. 
 This exact correspondence between the BFKL unintegrated
glue of the free nucleon \cite{BFKL} and nonlinear collective WW glue of the
nucleus in the calculation of these observables
 is a heartening finding. It persists despite the 
the sea quarks and antiquarks with
the transverse momenta below $Q_{A}$
being generated by the anticollinear, anti-DGLAP splitting of gluons
into sea, when the transverse momentum of the parent gluons is larger
than the momentum of the produced sea quarks. 

In \cite{Saturation} we noticed that less inclusive quantities 
like a spectrum of leading quarks from the truly inelastic DIS
or coherent diffractive breakup 
off nuclei are nonlinear functionals of the collective nuclear
WW glue.
Consequently, in the quest for factorization
properties of nuclear interactions one must go beyond the one-parton
observables such as the amplitude of
coherent diffractive breakup of pions or photons into dijets,
single-jet inclusive cross section and/or nuclear sea parton density. 
In this communication we discuss truly inelastic
hard interaction with nuclei followed by a breakup of the projectile
into forward hard dijets \footnote{The preliminary results from this study have
been reported elsewhere \cite{LIYaF,Conferences}}.
We illustrate our major point on an example of
DIS at small $x$ with a
breakup of the (virtual)
photon into a hard approximately back-to-back dijet with
small separation in rapidity, such that
the so-called lightcone plus-components
of the jet momenta sum up to the lightcone plus-component of
the photon's momentum, i.e., the so-called $x_{\gamma}=1$ criterion
is fulfilled (e.g., see \cite{ggFusion} and references
therein). In the familiar collinear approximation such
a dijet originates from the
photon-gluon fusion $\gamma^* g \to q\bar{q}$, often referred
to as an interaction of the unresolved or direct
photon. The allowance for the
transverse momentum of gluons leads to a disparity of the
momenta and to an azimuthal decorrelation of the quark and antiquark
jets which in DIS off free protons within the $k_{\perp}$-factorization
can be quantified
in terms of the unintegrated gluon SF of
the target (see \cite{Azimuth,Forshaw} and references therein).  
A substantial nuclear broadening
of the unintegrated gluon SF of nuclei at small $x$ and of the
nuclear sea  parton distributions \cite{Saturation,Mueller}
points at a stronger azimuthal decorrelation of jets produced in DIS
off nuclei. Furthermore, our finding of anticollinear, anti-DGLAP 
splitting of gluons into sea suggests
strongly a complete azimuthal decorrelation of forward quark and
antiquark jets with the transverse momenta below the saturation
scale, $p_{\pm} \lsim Q_A$. In this communication we
quantify these expectations and formulate a nonlinear 
generalization of the $k_{\perp}$-factorization
for inclusive dijet spectrum.

The technical basis of our approach is the color-dipole
multiple-scattering
theory of small-$x$  DIS off nuclei \cite{NZ91,NZZdiffr}.
We derive a consistent $k_{\perp}$-factorization
description of the azimuthal decorrelation of jets in terms of the
collective Weizs\"acker-Williams unintegrated gluon SF 
of the nucleus. In this derivation we follow closely
our early work \cite{Saturation} on the color-dipole
approach to saturation of nuclear partons. We focus on DIS at
$x\lsim x_A = 1/R_A m_N \ll 1$ which is dominated by interactions
of $q\bar{q}$ Fock states of the photon. Here $m_N$ is the nucleon
mass and $R_A$ is the radius 
of the target nucleus of mass number $A$. Nuclear attenuation of
these $q\bar{q}$ color dipoles \cite{NZ91,BGNPZshad} quantifies the fusion of
gluons and sea quarks from spatially overlapping nucleons of the
Lorentz contracted nucleus (\cite{NZfusion}, see also
\cite{Mueller1,McLerran}). 
Here we report also some of
the technical details, especially on the non-Abelian aspects of
propagation of color dipoles in nuclear matter, which were omitted
in the letter publication  \cite{Saturation}.

We focus on the genuinely inelastic DIS followed by color excitation
of the target nucleus. For heavy nuclei of equal importance is
coherent diffractive DIS in which the target nucleus does not break
and is retained in the ground state. Coherent diffractive DIS
 makes 50 per cent of the total DIS events at small $x$ 
\cite{NZZdiffr}, and in
these coherent diffractive events quark and antiquark jets are
produced exactly back-to-back with negligibly small transverse
decorrelation momentum $|\bDelta|=
|\bp_+ + \bp_-| \lsim 1/R_A \sim m_{\pi}/A^{1/3}$.

The further presentation is organized as follows. We work at the
parton level and discuss the transverse momentum distribution of
the final state quark and antiquark in interactions of $q\bar{q}$
Fock states of the photon with heavy nuclei. In section 2 we set
up
the formalism with a brief discussion of the decorrelation of jets
in DIS off free nucleons. In section 3 we report a derivation of
the general formula for the two-body transverse momentum
distribution. Color exchange between the initially color-neutral
$q\bar{q}$ dipole and nucleons of the target nucleus lead to an
intranuclear propagation of the color-octet $q\bar{q}$-states. Our
formalism, based on the technique \cite{NPZcharm,LPM}, includes
consistently the diffractive attenuation of color dipoles and
effects of transitions between color-singlet and color-octet
$q\bar{q}$-pairs, as well as between different color states of the
$q\bar{q}$-pair.  The hard jet-jet inclusive cross section is discussed in
section 4. For hard dijets diffractive attenuation effects are
weak and we obtain a nuclear $k_{\perp}$-factorizaton
 formula for the broadening of
azimuthal correlations between the quark and antiquark jets, which
is reminiscent of
that for a free nucleon target and is still a linear functional of
the collective WW gluon SF of the nucleus. We relate 
the decorrelation (acoplanarity) momentum to 
the nuclear saturation scale $Q_A$. In section 5 
working to large-$N_c$ approximation we
derive our central result -
a nonlinear nuclear $k_{\perp}$-factorization 
formula for the inclusive dijet
cross section and prove a complete disappearance of
the jet-jet correlation for minijets with the transverse momentum
below the saturation scale $Q_A$. 
In section 6 we present numerical estimates for the 
acoplanarity momentum distribution based on the unintegrated glue
of the proton determined in \cite{INDiffGlue}.
We point out a strong enhancement of decorrelations
from average to central DIS
and comment on possible relevance of our
mechanism of azimuthal decorrelations to the
recent observation of the dissolution of the away jets in central
nuclear collisions at RHIC \cite{RHIC_STAR}.
The next-to-leading order $1/N_c^2$ corrections to the large-$N_c$
results of section 5 are discussed in section 7. Here we derive the
nonlinear $k_{\perp}$-factorization representation for
the $\propto 1/N_c^2$ corrections and establish a close connection between
the $1/N_c^2$ and higher-twist expansions. 
In the Conclusions section we summarize
our principal findings.

Some of the technical details are presented in the Appendices.
In Appendix A we present the calculation of the matrix of 4-body
cross sections which enters the evolution operator for the
intranuclear propagation of color dipoles. In Appendix B we revisit 
the
single-jet spectrum and total cross section of DIS off
nuclei and demonstrate how the color-dipole extension \cite{NZ91,NZZdiffr} 
of the  Glauber-Gribov results
\cite{Glauber,Gribov} is recovered despite a nontrivial spectrum of
eigen-cross sections for non-Abelian propagation of color dipoles in a nuclear
matter. The properties of
collective unintegrated gluon SF for overlapping
nucleons of the Lorentz-contracted ultrarelativistic nucleus are
discussed in Appendix C.


\section{$k_{\perp}$-factorization for breakup of photons into
forward dijets in
DIS off free nucleons}

We recall briefly the color dipole formulation of DIS
\cite{NZ91,NZZdiffr,NZ92,NZ94,NZZlett} and set up a
formalism on an example of jet-jet decorrelation in DIS
off free nucleons which at moderately small $x$ is dominated by
interactions of $q\bar{q}$ states of the photon. The total cross
section for interaction of the color dipole $\br$ with the target
nucleon equals \cite{NZglue,BGNPZunit}
\bea
\sigma(r)&=& \alpha_S(r) \sigma_0\int d^2\bkappa
f(\bkappa )\left[1 -\exp(i\bkappa \br )\right]\nonumber\\
& = &
{1\over 2}\alpha_S(r) \sigma_0\int d^2\bkappa
f(\bkappa )\left[1 -\exp(i\bkappa \br )\right]
\cdot \left[1 -\exp(-i\bkappa \br )\right]
\, ,
\label{eq:2.1}
\eea
where $\sigma_0$ is an auxiliary soft parameter,
$f(\bkappa )$ is normalized as
$
\int d^2\bkappa  f(\bkappa )=1$ and is related to the BFKL
unintegrated gluon SF of
the target nucleon (\cite{BFKL}, for the phenomenology and review
see \cite{INDiffGlue,Andersson}) 
${\cal F}(x,\kappa^2) = {\partial G(x,\kappa^2)/
\partial\log\kappa^2}$ by
\bea
f(\bkappa ) = {4\pi \over
N_c\sigma_0}\cdot {1\over \kappa^4} \cdot {\cal F}(x,\kappa^2)\, .
\label{eq:2.2}
\eea

\begin{figure}[!t]
\begin{center}
\epsfig{file=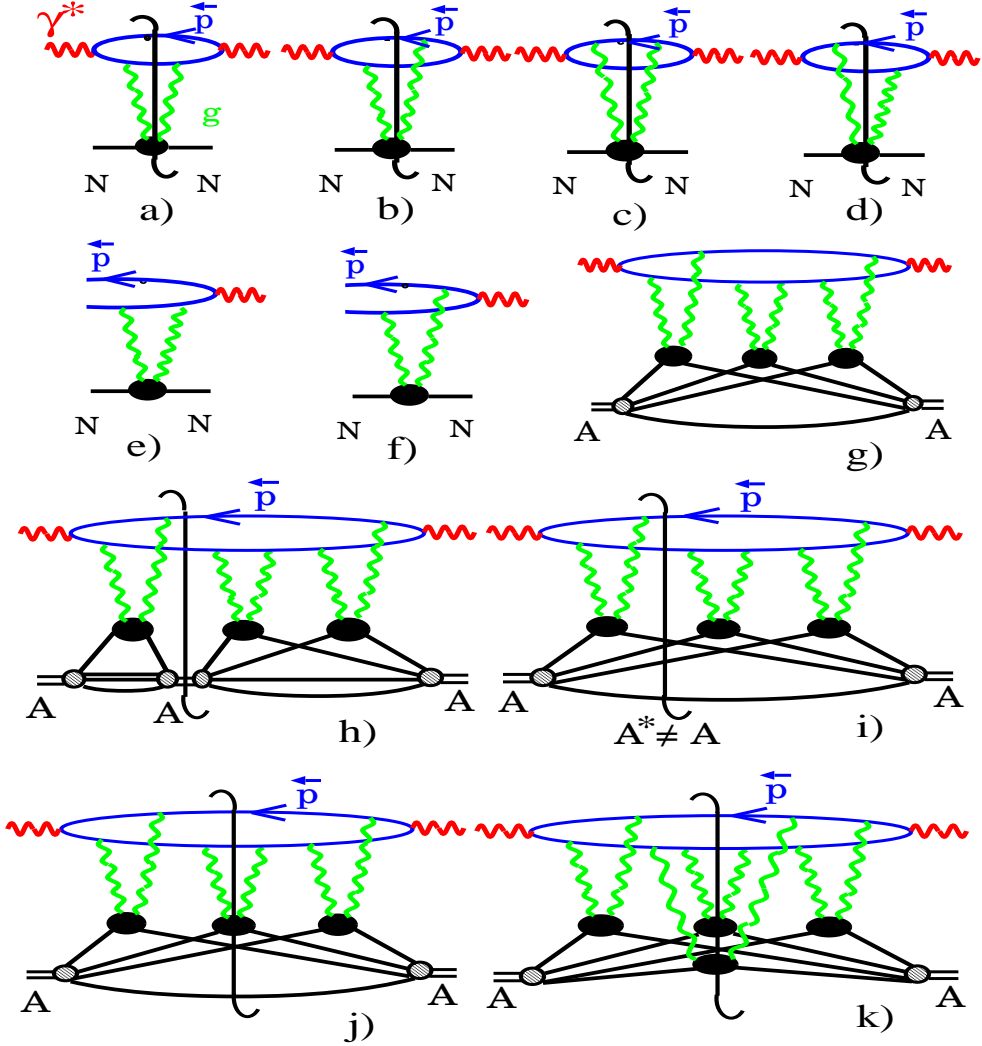, height=14.0cm, width = 13.0cm}
\end{center}
\caption{ \it The pQCD diagrams for cross section of inclusive DIS 
off nucleons (a-d) and nuclei
(g-k) 
and the amplitude of 
diffractive
 DIS off protons (e,f). Diagrams (a-d) show the unitarity cuts with color
excitation of the target nucleon, (g) - a generic multiple
scattering diagram for the amplitude of 
Compton scattering off nucleus, (h) - the
unitarity cut for a coherent diffractive DIS with retention of
the ground state nucleus $A$ in the final state, (i) - 
the unitarity
cut for quasielastic diffractive DIS with excitation and breakup 
of the
nucleus $A^*$, (j,k) - the unitarity cuts for truly inelastic DIS
with single (j) and multiple (k) 
color excitation of nucleons of the
nucleus. }
\end{figure}

For DIS off a free nucleon target, see figs. 1a-1d, the total
photoabsorption cross section equals \cite{NZ91}
\bea
\sigma_N(Q^2,x) = \int d^2\br dz |\Psi(Q^2,z,\br)|^2  \sigma(x,\br)\, ,
\label{eq:2.3}
\eea
where $\Psi(Q^2,z,\br)$ is the wave function of the $q\bar{q}$
Fock state of the photon, $Q^2$ and $x$ are the standard DIS variables. 
In the momentum 
representation,
\bea
{d\sigma_N \over d^2\bp_+ dz} =
{\sigma_0\over 2}\cdot { \alpha_S(\bp_+ ^2) \over (2\pi)^2}
 \int d^2\bkappa f(\bkappa )
\left|\langle \gamma^*|z,\bp_+\rangle - 
\langle \gamma^*|z,\bp_+ -\bkappa \rangle\right|^2\, ,
\label{eq:2.4}
\eea
where $\bp_+$ is the transverse momentum of the quark, the antiquark
has the transverse momentum $\bp_- = -\bp_+ + \bkappa$,  and $z_+= z$ and $
z_- = 1-z$
are the fractions of photon's lightcone momentum carried by the quark
and antiquark, respectively. The variables $z_{\pm}$ for the observed
jets add up to unity, $x_{\gamma}=z_+ + z_- = 1$, which in the realm of
DIS is referred to as the unresolved/direct photon interaction.

Upon summing over the helicities 
$\lambda,\overline{\lambda}$ of the
final state quark and antiquark for transverse photons
and flavor $f$ we have 
\bea
&&\left|\langle \gamma^*|z,\bp\rangle - \langle \gamma^*|z,\bp-\bkappa \rangle\right|^2_
{\lambda_{\gamma}=\pm 1} =
\nonumber\\
&&2N_c e_f^2\alpha_{em}\left\{
[z^{2}+(1-z)^{2}]
\left({\bp \over  \bp^{2}+\varepsilon^{2}} -
{\bp-\bkappa  \over  (\bp-\bkappa )^{2}+\varepsilon^{2}}
\right)^2_{\lambda+\overline{\lambda}=0}\right.\nonumber\\
&&\left. +m_{f}^{2}
\left({1   \over  \bp^{2}+\varepsilon^{2}}-
{1 \over  (\bp-\bkappa )^{2}+\varepsilon^{2}}\right)^2
_{\lambda+\overline{\lambda}=\lambda_{\gamma}}\right\}
\label{eq:2.5}
\eea
and for longitudinal photons
\bea
&&
\left|\langle \gamma^*|z,\bp\rangle - \langle \gamma^*|z,\bp-\bkappa \rangle\right|^2
_{\lambda_{\gamma}=0} = \nonumber\\
&&8N_c e_f^2\alpha_{em} Q^2z^2(1-z)^2
\left({1   \over  \bp^{2}+\varepsilon^{2}}-
{1 \over  (\bp-\bkappa )^{2}+\varepsilon^{2}}\right)^2 
_{\lambda+\overline{\lambda}=\lambda_{\gamma}}\, ,
\label{eq:2.6}
\eea
where
$\varepsilon^2 = z(1-z)Q^2 + m_f^2$.

Now, notice that the transverse momentum of the gluon is precisely the
decorrelation momentum $\bDelta=\bp_+ +\bp_-$, so that in the still further differential
from
\bea
{d\sigma_N \over dz d^2\bp_+ d^2\bDelta}& =&
{\sigma_0\over 2}\cdot { \alpha_S(\bp^2) \over (2\pi)^2}
 f(\bDelta )
\left|\langle \gamma^*|z,\bp_+\rangle -
\langle \gamma^*|z,\bp_+ -\bDelta \rangle\right|^2 \nonumber\\
& =& { \alpha_S(\bp^2) \over 2\pi N_c} \cdot 
 { {\cal F}(x,\bDelta^2) \over \Delta^4}
\cdot 
\left|\langle \gamma^*|z,\bp_+\rangle -
\langle \gamma^*|z,\bp_+ -\bDelta \rangle\right|^2\, .
\label{eq:2.7}
\eea

The small-$x$ result (\ref{eq:2.7}) shows that in DIS
forward dijets acquire their large transverse momentum from the
intrinsic momentum of quark and antiquark in the wave function of
the projectile photon, hence dubbing this process a breakup of the
photon into forward hard dijets is appropriate. 
Besides the criterion $x_{\gamma}=1$ the experimental signature of 
the photon breakup is a small rapidity separation of forward jets,
i.e., $z_+ \sim z_-$. 
The perturbative hard scale for our process is set by $(4\bp_+^2+Q^2)$ 
and gluon SF of the proton enters (\ref{eq:2.7}) at the Bjorken
variable $x=(4\bp_+^2+Q^2)/W^2$, where $W$ is the $\gamma^*p$ 
center of mass energy.
The purpose of our study is an
extension of (\ref{eq:2.7}) to breakup of photons into dijets
in truly inelastic DIS on nuclear targets.


\section{Breakup of photons into dijets on nuclear targets}

 We focus on DIS at
$x\lsim x_A = 1/R_A m_N \ll 1$ which is dominated by interactions
of $q\bar{q}$ states of the photon.  This is a starting
term of the leading $\log {1\over x}$ expansion, 
extension to interactions
of higher Fock states of the photon and corresponding
$\log {1\over x}$ evolution to smaller $x$  
will be discussed elsewhere. For $x\lsim x_A$ the
propagation of the $q\bar{q}$ pair inside nucleus can be treated
in the straight-path approximation.

We work in the conventional approximation of two t-channel gluons
in DIS off free nucleons. The relevant unitarity cuts of the
forward Compton scattering amplitude are shown in figs. 1a-1d
and describe the transition from the color-neutral $q\bar{q}$
dipole to color-octet $q\bar{q}$ pair
\footnote{To be more precise, for arbitrary $N_c$ color-excited
$q\bar{q}$ pair is in the adjoint representation  and
quarks in fundamental representation of $SU(N_c)$, our reference to the 
color octet and 
triplet must not cause any confusion.}. The two-gluon
exchange approximation amounts to neglecting unitarity 
constraints in DIS off free nucleons. As a quantitative
measure of
unitarity corrections one can take diffractive DIS off free
nucleons the amplitude of which 
 is described by higher order diagrams of fig. 1e,f
\cite{NZ92,NZ94,BGNPZunit} and which is only a 
small fraction of total DIS, $\eta_D \ll 1$
\cite{GNZ95,H1gap,ZEUSgap}. The unitarity cuts of 
the nuclear Compton scattering
amplitude which correspond to the genuine inelastic DIS with color
excitation of the nucleus are shown in figs. 1j,k. The 
diagram 1k describes a consecutive color excitation of 
the target nucleus accompanied by the color-space rotation
of the color-octet $q\bar{q}$.

Let $\bb_+$ and $\bb_-$ be the impact parameters of the quark
and antiquark, respectively, and $S_A(\bb_+,\bb_-)$  be the
S-matrix for interaction of the $q\bar{q}$ pair with the nucleus.
We are interested in the truly inelastic 
inclusive cross section summed
over all excitations of the target nucleus when one or several
nucleons have been color excited. A convenient way to sum such
cross sections is offered by the closure relation \cite{Glauber}. Regarding
the color states $c_{km}$ of the $q_k\bar{q}_m$ pair, we sum over all octet
and singlet states. Then the 2-jet inclusive
spectrum is calculated in terms of the 2-body density matrix as
\bea
&&{d\sigma_{in} \over dz d^2\bp_+ d^2\bp_-} =
{1\over (2\pi)^4} \int d^2 \bb_+' d^2\bb_-' d^2\bb_+ d^2\bb_- \nonumber\\
&&\times \exp[-i\bp_+(\bb_+ -\bb_+')-i\bp_-(\bb_- -
\bb_-')]\Psi^*(Q^2,z,\bb_+' -\bb_-')
\Psi(Q^2,z,\bb_+ -\bb_-)\nonumber\\
&&\times \Bigl\{ \sum_{A^*} \sum_{km}  \langle
1;A|S_A^*(\bb_+',\bb_-')|A^*;c_{km}\rangle \langle
c_{km};A^*|S_A(\bb_+,\bb_-)|A;1\rangle  \nonumber\\
&& -
\langle 1;A|S_A^*(\bb_+',\bb_-')
|A;1\rangle
\langle 1;A|S_A(\bb_+,\bb_-)|A;1\rangle \Bigr\}\, .
\label{eq:3.1}
\eea
 In the
integrand of (\ref{eq:3.1}) we subtracted the coherent 
diffractive component
of the final state. Notice, that four straight-path trajectories
$\bb_{\pm},\bb'_{\pm}$ enter the calculation of the full fledged
2-body density matrix  and $S_A$ and $S_A^*$ describe the propagation of
two quark-antiquarks pairs, $q\bar{q}$ and $q'\bar{q}'$, 
inside a nucleus.

The further analysis of the integrand of (\ref{eq:3.1}) is
a non-Abelian generalization of the formalism 
developed by one of the authors (BGZ)
for the in-medium evolution of ultrarelativistic positronium
\cite{BGZpositronium}.  
Upon the application of closure to sum over nuclear final states
$A^*$ the integrand of (\ref{eq:3.1}) can be considered as an
intranuclear evolution operator for the 2-body density matrix
(for the related
discussion see also ref. \cite{NSZdist}) 
 \bea \sum_{A^*}
\sum_{km} \langle A| \Bigl\{ \langle 1|
S_A^*(\bb_+',\bb_-')|c_{km}\rangle \Bigr\} |A^* \rangle \langle
A^*| \Bigl\{ \langle c_{km}| S_A(\bb_+,\bb_-)|1\rangle \Bigr\}
|A\rangle =\nonumber\\
=\langle A| \left\{ \sum_{km} \langle 1|
S_A^*(\bb_+',\bb_-')|c_{km}\rangle \langle
c_{km}|S_A(\bb_+,\bb_-)|1\rangle\right\} |A\rangle \, .
\label{eq:3.2}
\eea 
Let the eikonal for the quark-nucleon and
antiquark-nucleon QCD gluon exchange interaction be $T^a_{+}
\chi(\bb)$ and  $T^a_{-} \chi(\bb)$, where  $T^{a}_+$ and
$T^{a}_{-}$ are the $SU(N_c)$ generators for the quark and
antiquarks states, respectively. The vertex $V_a$ for excitation
of the nucleon $g^a N \to N^*_a$ into color octet state is so
normalized that after application of closure the vertex $g^a g^b
NN$ in the diagrams of fig. 1a-d is $\delta_{ab}$. Then, to the
two-gluon exchange approximation, the $S$-matrix of the
$(q\bar{q})$-nucleon interaction equals 
\bea 
S_N(\bb_+,\bb_-)  = 1
+ i[T^a_{+} \chi(\bb_+)+ T^a_{-} \chi(\bb_-)]V_a - {1\over 2}
[T^a_{+} \chi(\bb_+)+ T^a_{-} \chi(\bb_-)]^2 \, .
\label{eq:3.3}
\eea
The profile function for interaction of the $q\bar{q}$ dipole
with a nucleon is 
$\Gamma(\bb_+,\bb_-)= 1 - S_N(\bb_+,\bb_-)$.
For a color-singlet dipole 
$(T^a_{+}+ T^a_{-})^2=0$ and 
the dipole cross section for interaction of the color-singlet
$q\bar{q}$ dipole with the nucleon equals
\bea
\sigma(\bb_+-\bb_-) = 2\int d^2\bb_{+} \langle N|\Gamma(\bb_+,\bb_-)
|N\rangle =
{N_c^2 -1 \over 2 N_c} \int
d^2\bb_+ [\chi(\bb_+)-\chi(\bb_-)]^2\, . 
\label{eq:3.4} \eea
The nuclear $S$-matrix of the straight-path approximation is
$$
S_A(\bb_+,\bb_-) = \prod _{j=1}^A S_N(\bb_+ -\bb_j,\bb_- - \bb_j)\, ,
$$
where the ordering along the longitudinal path is understood.
We evaluate the nuclear expectation value in (\ref{eq:3.2})
in the standard dilute gas approximation.
To the two-gluon exchange approximation, per each and every nucleon
$N_j$ only the terms quadratic
in $\chi(\bb_j)$ must be kept in the evaluation of the
single-nucleon matrix element
$$
\langle N_j|S_N^*(\bb_+' -\bb_j,\bb_-' - \bb_j)
S_N(\bb_+-\bb_j,\bb_- - \bb_j)|N_j \rangle
$$
 which enters the calculation of
$S_A^*S_A$. Following the technique developed in
\cite{NPZcharm,LPM} we can reduce the calculation of the evolution
operator for the 2-body density matrix (\ref{eq:3.2}) to the
evaluation of the $S$-matrix $S_{4A}(\bb_+,\bb_-,\bb_+',\bb_-')$
for the scattering of a fictitious 4-parton state composed of the
two quark-antiquark pairs in the overall color-singlet state.
Because $(T^{a}_+)^*= -T^{a}_{-}$, within the two-gluon exchange
approximation the quarks entering the complex-conjugate $S_A^*$ in
(\ref{eq:3.2}) can be viewed as antiquarks, so that 
\bea
&&\sum_{km} \langle 1| S_A^*(\bb_+',\bb_-')|c_{km}\rangle \langle
c_{km}|S_A(\bb_+,\bb_-)|1\rangle \nonumber\\
&&=\sum_{km jl} \delta_{kl} \delta_{mj}
\langle c_{km}c_{jl}|S_{4A}(\bb_+',\bb_-',\bb_+,\bb_-)|11\rangle \, ,
\label{eq:3.5}
\eea
where $S_{4A}(\bb_+',\bb_-',\bb_+,\bb_-)$ is an
S-matrix for the propagation of the two quark-antiquark pairs in the
overall singlet state. While the first $q\bar{q}$ pair is formed
by the initial quark $q$ and antiquark $\bar{q}$ at impact parameters
$\bb_{+}$ and $\bb_-$, respectively, in the second $q'\bar{q}'$ pair 
the quark $q'$ propagates at an impact parameter $\bb_-'$ and the
antiquark $\bar{q}'$ at an impact parameter $\bb_+'$.
In the initial state  both quark-antiquark pairs are in
color-singlet states: $ | in \rangle = |11\rangle$.

Let us introduce the normalized singlet-singlet and octet-octet
states
\beq
|11\rangle = {1\over N_c} (\bar{q} q)(\bar{q}' q')\, , ~~
|88\rangle = {2\over \sqrt{N_c^2 -1}} (\bar{q}T^a q)(\bar{q}' T^a q')\, ,
\label{eq:3.6}
\eeq
where $N_c$ is the number of colors and $T^{a}$ are the generators
of $SU(N_c)$. Making use of the color Fiertz identity,
\beq
\delta^k_j \delta_l^m = {1\over N_c} \delta^k_l \delta^m_j +
2 \sum _{a} (T^{a})^k_l(T^{a})^m_j \, ,
\label{eq:3.7}
\eeq
the sum (\ref{eq:3.5}) over
color states of the produced quark-antiquark pair can be
represented as
\bea
\sum_{km}
\langle c_{km} c_{km}|S_{4A}(\bb_+',\bb_-',\bb_+,\bb_-)|11\rangle =
\langle 11|S_{4A}(\bb_+',\bb_-',\bb_+,\bb_-)|11\rangle \nonumber\\
+ \sqrt{N_c^2 -1} \langle 88|S_{4A}(\bb_+',\bb_-',\bb_+,\bb_-)|11\rangle
\, .
\label{eq:3.8}
\eea

If $\sigma_4(\bb_+',\bb_-',\bb_+,\bb_-)$ is the color-dipole cross
section operator for the 4-body state, then the 
evaluation of the nuclear expectation value for a dilute gas
nucleus in the standard approximation of
neglecting the size of color dipoles compared to a radius of heavy
nucleus gives
\cite{Glauber}
 \bea S_{4A}(\bb_+',\bb_-',\bb_+,\bb_-)=\exp\{-
{1\over 2}\sigma_{4}(\bb_+',\bb_-',\bb_+,\bb_-)T(\bb)\}\, ,
\label{eq:3.9} 
\eea 
where 
$ T(\bb)=\int db_z n_{A}(b_z, \bb)$ 
is the optical thickness of a nucleus at an
impact parameter 
$
\bb = {1\over 4}( \bb_+ + \bb_+' + \bb_- + \bb_-')\, ,
$
(one should not confuse $\bb$ with the center of gravity of
color dipoles where the impact parameters $\bb_{\pm}$ and $\bb_{\pm}'$
must be weighted with $z_{\pm}$, the difference between the two
quantities is irrelevant here) and $ n_{A}(b_z, \bb)$
is nuclear matter density with the normalization 
$\int d^2\bb T(\bb) = A$ .  The single-nucleon $S$-matrix (\ref{eq:3.3}) contains
transitions from the color-singlet to the both color-singlet and
color-octet $q\bar{q}$ pairs. However, only  color-singlet
operators contribute to $\langle N_j|S_N^*(\bb_+' -\bb_j,\bb_-' -
\bb_j) S_N(\bb_+-\bb_j,\bb_- - \bb_j)|N_j \rangle$, and hence
the matrix
$\sigma_4(\bb_+',\bb_-',\bb_+,\bb_-)$ only includes transitions
between the $|11\rangle$ and $|88\rangle$ color-singlet 4-parton
states, the $|18\rangle$ states are not allowed.

\begin{figure}[!t]
   \centering
   \epsfig{file=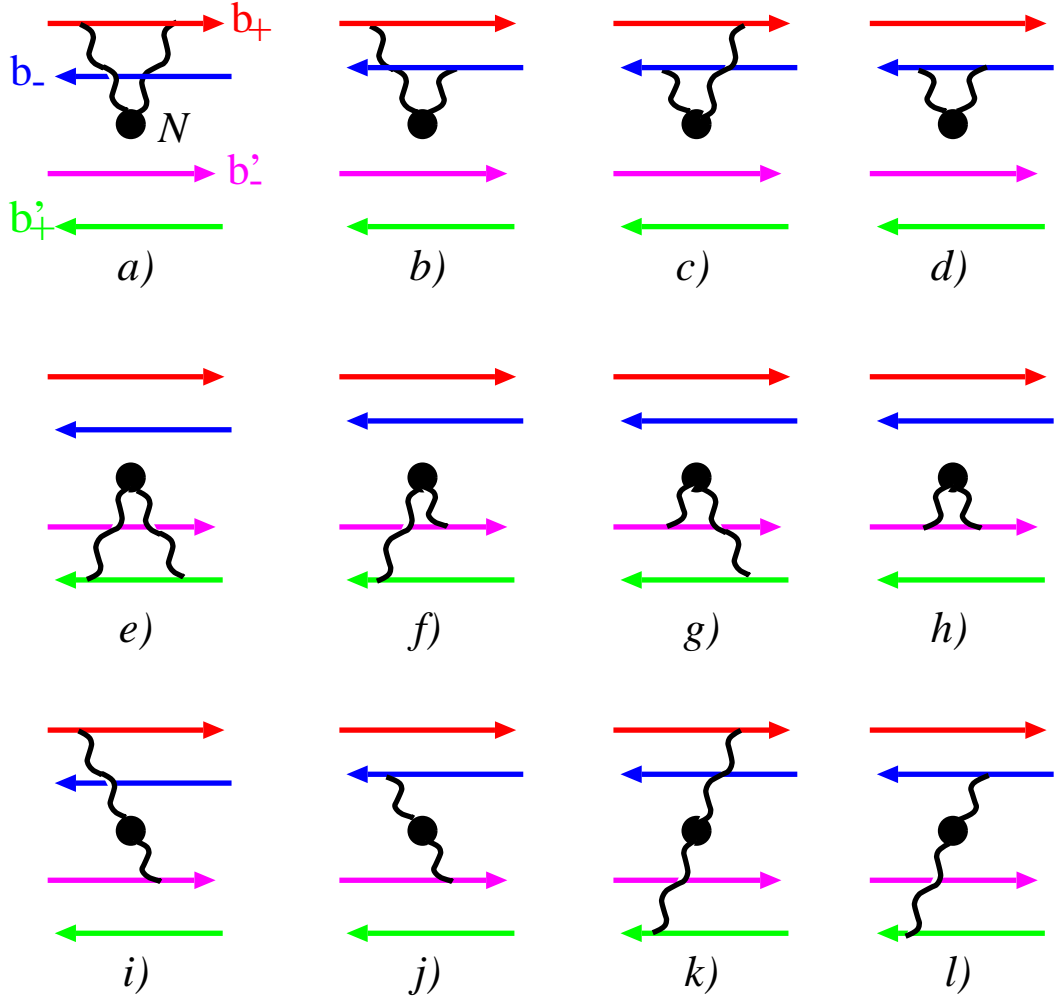,width=14cm}
\caption{\it The pQCD diagrams for the matrix of color dipole
cross section for the
4-body $(q\bar{q})(q'\bar{q}')$ state. The sets 4a-4d and
4e-4h show the diagrams for the scattering without
changing the color state of the $q\bar{q}$ and $q'\bar{q}'$
dipoles, the set 4i-4l shows only half of
the diagrams for scattering with rotation of the color state of dipoles.}
\end{figure}

The pQCD diagrams for the 4-body cross section are shown in fig.2.
 It is convenient to
introduce 
\beq
\bs = \bb_+ - \bb_+' \, ,
\label{eq:3.10}
\eeq
for the variable conjugate to the decorrelation momentum,
and $\br=\bb_+ - \bb_-,~\br'= \bb_+'- \bb_-'$, 
in terms of which
\bea
\bb_+ - \bb_-' = \bs + \br'\, ,~~
\bb_- - \bb_+' = \bs - \br \, ,~~
\bb_- - \bb_-' = \bs -\br + \br'\, .
\label{eq:3.11}
\eea
Performing the relevant color
algebra, we find (some details of the derivation are presented in
Appendix A)
\bea 
\sigma_{11}=\langle 11|\sigma_4|11\rangle = \sigma(\br)+
\sigma(\br') \, ,
\label{eq:3.12} 
\eea 
\bea 
\sigma_{18}=
\langle
11|\sigma_4|88\rangle &=& {\sigma(\bs)+\sigma(\bs -\br + \br')-
 \sigma(\bs + \br')- \sigma(\bs - \br) \over \sqrt{N_c^2-1}} \nonumber\\
&= & -{\Sigma_{18}(\bs,\br,\br') \over \sqrt{N_c^2-1}} \, ,
\label{eq:3.13}
\eea
\bea
\sigma_{88}=\langle 88|\sigma_4|88\rangle =
{N_c^2 -2 \over N_c^2-1}
[\sigma(\bs)+\sigma(\bs -\br + \br')]\nonumber\\
+{2 \over N_c^2-1}[\sigma(\bs + \br')+ \sigma(\bs - \br)]
-{1\over N_c^2-1}
[\sigma(\br)+\sigma(\br')] \, .
\label{eq:3.14}
\eea
The term in (\ref{eq:3.1}), which subtracts the contribution from 
diffractive processes
without color excitation of the target nucleus, equals
\bea
&&\langle 1;A|S_A^*(\bb_+',\bb_-')
|A;1\rangle
\langle
1;A|
S_A(\bb_+,\bb_-)|A;1\rangle \nonumber\\
&&= \exp\left\{- {1\over 2}\left[\sigma(\br)+\sigma(\br')\right]T(\bb)
\right\}=\exp \left\{-{1\over 2}\sigma_{11}
T(\bb)\right\} \, .
\label{eq:3.15}
\eea

In the discussion of nuclear effects 
it is convenient to use the Sylvester expansion
\bea \exp\{- {1\over 2}\sigma_4 T(\bb)\} = \exp\{- {1\over
2}\Sigma_1 T(\bb)\}{\sigma_4 - \Sigma_2 \over \Sigma_1 - \Sigma_2}
+  \exp\{- {1\over 2}\Sigma_2 T(\bb)\}{\sigma_4 - \Sigma_1 \over
\Sigma_2 - \Sigma_1}  \, ,
\label{eq:3.17} 
\eea 
where 
$\Sigma_{1,2}$ are the two eigenvalues of the operator
$\sigma_4$,
\beq
\Sigma_{1,2} = {1\over 2} (\sigma_{11}+\sigma_{88}) \mp
{1\over 2}(\sigma_{11}-\sigma_{88}) 
\sqrt{1 + {4\sigma_{18}^2 \over (\sigma_{11}-\sigma_{88})^2}} \, .
\label{eq:3.16}
\eeq
An
application to (\ref{eq:3.8}) of the Sylvester expansion gives
for the integrand of (\ref{eq:3.1}) 
\bea &&\sum_{A^*} \sum_{km}
\langle 1;A|S_A^*(\bb_+',\bb_-')|A^*;c_{km}\rangle \langle
c_{km};A^*|S_A(\bb_+,\bb_-)|A;1\rangle  \nonumber\\
&& - \langle 1;A|S_A^*(\bb_+',\bb_-') |A;1\rangle
\langle 1;A|S_A(\bb_+,\bb_-)|A;1\rangle  
\nonumber\\
&=&(\langle 11| + \sqrt{N_c^2 -1} \langle 88|) \exp\left\{-{1\over
2}\sigma_4 T(\bb)\right\}|11\rangle - 
\exp\left\{-{1\over2}\sigma_{11}
T(\bb)\right\}\nonumber\\
&=& \exp\left\{-{1\over 2}\Sigma_2 T(\bb)\right\}-
\exp\left\{-{1\over 2}\sigma_{11}
T(\bb)\right\}\nonumber\\
&+& { \sigma_{11} - \Sigma_2 \over \Sigma_1 -
\Sigma_2} \left\{\exp\left[-{1\over 2}\Sigma_1 T(\bb)\right]-
\exp\left[-{1\over 2}\Sigma_2 T(\bb)\right]\right\}\nonumber\\
&+&{\sqrt{N_c^2 -1} \sigma_{18} \over
\Sigma_1 - \Sigma_2} \left\{\exp\left[-{1\over 2}\Sigma_1
T(\bb)\right]- \exp\left[-{1\over 2}\Sigma_2 T(\bb)\right]\right\} 
\, .
\label{eq:3.18} \eea


\section{Breaking of photons into hard dijets: a still linear 
nuclear $k_{\perp}$-
factorization }

The diagonalization of the $2\times 2$ matrix $\sigma_4$ is a
straightforward task, so that technically eqs. (\ref{eq:3.1}) and
(\ref{eq:3.18}) allow a direct calculation of the
jet-jet inclusive cross section in terms of the color dipole cross
section $\sigma(\br)$. The evaluation of the 6-fold Fourier
transform is not a trivial task, though.

First, notice that the difference between $\Sigma_2$ and 
$\sigma_{11}=\sigma(\br)+\sigma(\br')$ is quadratic or higher
order in the off-diagonal $\sigma_{18}$, see eq.~(\ref{eq:3.16}).
Consequently, the first two lines in the Sylvester expansion (\ref{eq:3.18})
start with terms $\propto \sigma_{18}^2$, whereas the last
line starts with terms $\propto \sigma_{18}$. Then 
it is convenient to represent (\ref{eq:3.18}) as
the impulse approximation (IA) term times the nuclear distortion
factor $D_A(\bs,\br,\br',\bb)$,
\bea
 &&\sum_{A^*} \sum_{km}
 \langle
1;A|S_A^*(\bb_+',\bb_-')|A^*;c_{km}\rangle \langle
c_{km};A^*|S_A(\bb_+,\bb_-)|A;1\rangle
\nonumber\\
&-&
 \langle 1;A|S_A^*(\bb_+',\bb_-') |A;1\rangle
\langle 1;A|S_A(\bb_+,\bb_-)|A;1\rangle  \nonumber\\
&=& T(\bb)\Sigma_{18}(\bs,\br,\br')
D_A(\bs,\br,\br',\bb)\, ,
 \label{eq:4.1} \eea 
so that 
\bea {d\sigma_{in}
\over d^2\bb dz d^2\bp_+ d^2\bp_-} &=& {1\over 2(2\pi)^4} \int
d^2\bs d^2\br d^2\br'\exp[-i(\bp_+ +\bp_-)\bs +i\bp_-(\br' -\br)]
\nonumber\\
&\times& \Psi^*(Q^2,z,\br')
\Psi(Q^2,z,\br) T(\bb)
\Sigma_{18}(\bs,\br,\br')
D_A(\bs,\br,\br',\bb) \, .
\label{eq:4.2}
\eea

As an introduction to nuclear $k_{\perp}$-factorization, 
we start with forward hard
jets with the momenta $\bp_{\pm}^2 \gsim Q_A^2$,
which  are produced from
interactions with the target nucleus of small color dipoles
in the incident photon such that diffractive nuclear attenuation
effects can be neglected. 
We proceed with the formulation of 
the Fourier representations for each  factor
in (\ref{eq:4.1}). The application of the integral
representation (\ref{eq:2.1}) gives : 
\bea 
&&\Sigma_{18}(\bs,\br,\br')=
\left[ \sigma(\bs)-
\sigma(\bs + \br')- \sigma(\bs -\br) + \sigma(\bs - \br + \br')
\right]
\nonumber\\
&=& \alpha_S \sigma_{0} \int d^2\bkappa f(\bkappa)
\exp[i\bkappa\bs] \left\{1 - \exp[i\bkappa \br']\right\} \left\{1
- \exp[-i\bkappa \br]\right\}\, . 
\label{eq:4.4} 
\eea 
Hard jets correspond to $|\br|, |\br'| \ll |\bs|$.
Then the two eigenvalues are
$
\Sigma_2\approx \sigma_{11}
$
and
$
\Sigma_1\approx  \sigma_{88} \approx
 2\lambda_c \sigma(\bs)
$, where $\lambda_c = N_c^2/(N_c^2-1) = C_A/2C_F$,
where
$C_F$ and $C_A$ are the Casimir operators for the 
fundamental and adjoint representations of
$SU(N_c)$. Because of $
\Sigma_2\approx \sigma_{11} \approx 0
$
only the last term,
 $\propto \sigma_{18} $, must be kept in the
Sylvester expansion (\ref{eq:3.18}), and 
the nuclear
distortion factor takes on a simple form
\bea
D_A(\bs,\br,\br',\bb)& =&  { 2 \over (\Sigma_2 - \Sigma_1)T(\bb)}
\left\{\exp\left[-{1\over 2}\Sigma_1 T(\bb)\right]-
\exp\left[-{1\over 2}\Sigma_2 T(\bb)\right]\right\}\nonumber\\
& =& {1-\exp \left[-{1\over 2}\Sigma_1 T(\bb)\right]
\over {1\over 2}\Sigma_1 T(\bb) } \, .
\label{eq:4.3}
\eea
The Fourier representation for the nuclear distortion factor
$D_A(\bs,\br,\br')$ is readily obtained making use
of the NSS
representation for the nuclear attenuation factor
\cite{Saturation,NSSdijet} 
\bea \exp\left[-{1\over 2}\sigma(\bs)
T(\bb)\right] &=& \exp\left[-\nu_{A}(\bb)\right]
\exp\left[\nu_{A}(\bb)\int
d^2\bkappa f(\bkappa)\exp(i\bkappa\bs)\right]\nonumber\\
&=&\exp\left[-\nu_{A}(\bb)\right] \sum_{j=0}^{\infty}
{\nu_{A}^{j}(\bb) \over j!}\int d^2\bkappa
f^{(j)}(\bkappa)\exp(i\bkappa\bs) \nonumber\\
&=& \int d^2\bkappa
\Phi(\nu_{A}(\bb), \bkappa) \exp(i\bkappa\bs)  \, .
\label{eq:4.5} 
\eea
in terms
of the nuclear Weizs\"acker-Williams glue per unit area in the
impact parameter plane, $\phi_{WW}(\nu_{A}(\bb),\bkappa)$, as defined in
\cite{Saturation}, 
\bea 
\Phi(\nu_{A}(\bb), \bkappa)= \sum_{j=0} w_{j}(\nu_A(\bb))f^{(j)}(\bkappa) =
\exp(-\nu_{A}(\bb))f^{(0)}(\bkappa) +\phi_{WW}(\nu_{A}(\bb),\bkappa)\, .
\label{eq:4.6} 
\eea 
Here 
\bea
\nu_{A}(\bb)= {1\over 2}\alpha_S(r)\sigma_0 T(\bb) 
\label{eq:4.6*}
\eea
and 
\bea
w_{j}(\nu_A(\bb))=
{\nu_{A}^{j}(\bb) \over j!}\exp\left[-\nu_{A}(\bb)\right]
\label{eq:4.7}
\eea
is a probability of finding $j$ spatially overlapping nucleons 
in a Lorentz-contracted 
nucleus, 
\beq 
f^{(j)}(\bkappa )= \int \prod_{i=1}^j
d^2\bkappa _{i} f(\bkappa _{i}) \delta(\bkappa -\sum_{i=1}^j
\bkappa _i) \,, ~~f^{(0)}(\bkappa)=\delta(\bkappa) 
\label{eq:4.8} 
\eeq 
is a collective gluon field of $j$ overlapping nucleons. As
usual, the strong coupling in (\ref{eq:4.6*}) must be taken
at the hardest relevant scale \cite{Dokshitser}. 

The denominator $\Sigma_1$ in
(\ref{eq:4.3}) is problematic from the point of view of the
Fourier transform but can
be eliminated by the integral representation
\bea
D_A(\bs) =
\int_0^1 d \beta \exp\left[-{1\over 2}\beta \Sigma_1 T(\bb)\right] 
= \int_0^1 d \beta  \int d^2\bkappa
\Phi(2\beta \lambda_c  \nu_A(\bb),\bkappa)\exp(i\bkappa \bs)\,.
\label{eq:4.9}
\eea
Here $\beta$ has a meaning of the fraction of the nuclear thickness
which the $(q\bar{q})$
pair propagates in the color octet state. 
The introduction of this distortion factor into (\ref{eq:4.2}) is straightforward
and gives our central result for the hard jet-jet inclusive cross section:
\bea
{d\sigma_{in} \over d^2\bb dz d^2\bp_+ d^2\bDelta}= T(\bb) \int d^2\bkappa
\int_0^1 d \beta \Phi(2\beta \lambda_c  \nu_A(\bb),\bDelta - \bkappa)
{d\sigma_{N} \over dz d^2\bp_+ d^2\bkappa }\, .
\label{eq:4.10}
\eea
Since for hard jets $\br^2 \sim 1/\bp_+^2$, one must use 
$\alpha_S(\bp_+^2)$ in the evaluation of $\nu_A(\bb)$.  
For a thin nucleus such that $\nu_A(\bb) \ll 1$, we have
$\Phi(2\beta \lambda_c  \nu_A(\bb),\bDelta - \bkappa) = \delta(\bDelta - \bkappa)$,
see eq.~(\ref{eq:4.6}),
and recover the IA result
\bea
{d\sigma_{in} \over d^2\bb dz d^2\bp_+ d^2\bDelta}= T(\bb)
{d\sigma_{N} \over dz d^2\bp_+ d^2\bDelta }\, .
\label{eq:4.11}
\eea
Our result
(\ref{eq:4.10}) for nuclear  broadening of
the acoplanarity momentum distribution of hard dijets can be regarded
as a nuclear counterpart of the $k_{\perp}$-factorization result (\ref{eq:2.7})
for the
free nucleon target.

The probabilistic form of a 
convolution eq.~(\ref{eq:4.10}) of the differential cross
section on a free nucleon target with the manifestly positive 
defined distribution $\Phi(2\beta \lambda_c  \nu_A(\bb),\bkappa)$
can be
understood as follows. Hard jets
originate from small color dipoles. Their interaction with gluons
of the target nucleus is suppressed by 
the mutual neutralization of color charges
of the quark and antiquark in the small-sized 
color-singlet $q\bar{q}$ state which
is manifest from the small cross section for a free nucleon target,
see eq. (\ref{eq:2.7}). The first inelastic
interaction inside a nucleus converts the $q\bar{q}$ pair into
the color-octet state, in which color charges of the quark and 
antiquark do not neutralize each other, rescatterings of the quark
and antiquark in the collective color field of intranuclear nucleons become
uncorrelated, and the broadening of the momentum
distribution with nuclear thickness follows a probabilistic
picture.


\section{Nonlinear nuclear $k_{\perp}$-factorization for breakup 
of photons into semihard
dijets: large-$N_c$ approximation}

Now we are in the position to relax the hardness restriction and
consider the semihard dijets, $|\bp_{\pm}| \sim Q_A$. In this section
we give a consistent treatment of this case in the venerable
large-$N_c$ approximation.
What we shall
formulate can be dubbed the nonlinear nuclear generalization of the
$k_{\perp}$-factorization.

The crucial point is that in the large-$N_c$ approximation
$\Sigma_{2} = \sigma_{11} = \sigma(\br)+\sigma(\br')$, so that
only the
last term in the Sylvester expansion (\ref{eq:3.18}) contributes to the
jet-jet inclusive cross section.
The nuclear distortion factor will
be still given by eq.~(\ref{eq:4.3}) but for finite $\Sigma_2$.
Upon the slight generalization of (\ref{eq:4.9}) and making use
of
\bea
\Sigma_1 = \sigma(\bs)+\sigma(\bs+\br'-\br)
\label{eq:5.1}
\eea
the distortion factor
can be cast in the form
\bea
D_A(\bs,\br,\br',\bb) &=&
\int_0^1 d \beta \exp\left\{-{1\over 2}[\beta \Sigma_1 + (1-\beta)\Sigma_2]
T(\bb)\right\}\nonumber\\
&=&
\int_0^1 d \beta \exp\left\{-{1\over 2}(1-\beta)[\sigma(\br)+\sigma(\br')]
T(\bb)\right\}  \nonumber\\
&\times&\exp\left\{-{1\over 2}\beta[\sigma(\bs)+\sigma(\bs+\br'-\br)]
T(\bb)\right\}
\label{eq:5.2}
\eea
the different exponential factors in which
admit a simple interpretation: The former two exponential factors describe
the intranuclear distortion of the incoming color-singlet $(q\bar{q}) \&
(q'\bar{q}')$
dipole state, whereas the last two factors describe the distortion of
the outgoing color-octet $(q\bar{q})\&
(q'\bar{q}')$ states. Application of the 
NSS representation to attenuation factors in (\ref{eq:5.2}) yields
\bea
D_A(\bs,\br,\br',\bb)= \int_0^1 d \beta && \int d^2\bkappa_1
\Phi((1-\beta)\nu_A(\bb),\bkappa_1)\exp(-i\bkappa_1 \br)
\nonumber\\
\times &&\int d^2\bkappa_2
\Phi((1-\beta)\nu_A(\bb),\bkappa_2)\exp(i\bkappa_2 \br)
\nonumber\\
\times &&\int d^2\bkappa_3
\Phi(\beta\nu_A(\bb),\bkappa_3)\exp[i\bkappa_3(\bs+\br'- \br)]
\nonumber\\
\times &&\int d^2\bkappa_4
\Phi(\beta\nu_A(\bb),\bkappa_4)\exp(i\bkappa_4 \br) \label{eq:5.3}
\eea The integral representation (\ref{eq:5.2}) furnishes two
important tasks: it removes $\Sigma_1 -\Sigma_2$ form the
denominator in (\ref{eq:3.18}) and gives for the nuclear
distortion factor the Fourier transform (\ref{eq:5.3}) which
is a product of the manifestly positive defined nuclear WW gluon
distributions.
Finally, the jet-jet inclusive cross section takes the form 
\bea
{d\sigma_{in} \over d^2\bb dz d\bp_{-} d\bDelta} &=& {1\over
2(2\pi)^2} \alpha_S \sigma_0 T(\bb)\int_0^1 d \beta \int
d^2\bkappa_1 d^2\bkappa_2 d^2\bkappa_3 d^2\bkappa f(\bkappa)
\nonumber\\
&\times & \Phi(\beta\nu_A(\bb),\bDelta -\bkappa_3 -\bkappa)
\Phi(\beta\nu_A(\bb),\bkappa_3)
\nonumber\\
&\times &\Phi((1-\beta)\nu_A(\bb),\bkappa_1)
\Phi((1-\beta)\nu_A(\bb),\bkappa_2)
\nonumber\\
&\times &\left\{\langle \gamma^*|z,\bp_{-} +\bkappa_2 +\bkappa_3\rangle
-
\langle \gamma^*|z,\bp_{-} +\bkappa_2 +\bkappa_3+\bkappa\rangle\right\}\nonumber\\
&\times &
\left\{\langle z,\bp_{-} +\bkappa_1 +\bkappa_3|\gamma^*\rangle -
\langle z,\bp_{-} +\bkappa_1 +\bkappa_3+\bkappa|\gamma^*\rangle\right\}
\nonumber\\
&=& {1\over
2(2\pi)^2} \alpha_S \sigma_0 T(\bb)\int_0^1 d \beta \int
d^2\bkappa_3 d^2\bkappa f(\bkappa)
\nonumber\\
&\times & \Phi(\beta\nu_A(\bb),\bDelta -\bkappa_3 -\bkappa)
\Phi(\beta\nu_A(\bb),\bkappa_3)
\nonumber\\
&\times &
\Biggl|\int d^2\bkappa_1\Phi((1-\beta)\nu_A(\bb),\bkappa_1)\nonumber\\
&\times&
\left\{\langle \gamma^*|z,\bp_{-} +\bkappa_1 +\bkappa_3\rangle -
\langle \gamma^*| z,\bp_{-} +\bkappa_1 +\bkappa_3+\bkappa\rangle\right\}\Biggr|^2
\, .
\label{eq:5.4}
\eea
This is our central result for inclusive cross
section of the photon breakup into dijets off nuclei.
It demonstrates how the broadening of
the transverse momentum distribution of dijets is uniquely
calculable in terms of the collective WW glue of a nucleus and as
such must be regarded as a nonlinear $k_{\perp}$-factorization for inclusive
dijet cross section. 

The last form of (\ref{eq:5.4}) shows clearly
that the integrand is manifestly positive valued. Going back to 
(\ref{eq:5.2}) and (\ref{eq:5.3}) one can identify the convolution
of the collective nuclear WW glue $\Phi((1-\beta)\nu_A(\bb),\bkappa_1)$
with the photon wave functions in the last form of (\ref{eq:5.4})
as an effect of distortions of the photon wave function when
$q\bar{q}$ pair was propagating in still the color-singlet state. 

Finally, consider the limiting case of $|\bp_-|,|\bDelta| \lsim Q_A$.
In our analysis \cite{Saturation} of the single particle spectrum we
discovered that the transverse momentum distribution of sea quarks
is dominated by the anticollinear, anti-DGLAP splitting of gluons
into sea, when the transverse momentum of the parent gluons is larger
than the momentum of the sea quarks. As we stated in the Introduction,
that suggests strongly
a complete azimuthal decorrelation of forward minijets with the transverse
momenta below the saturation scale, $p_{\pm} \lsim Q_A$. Our analysis of 
$f^{(j)}(\bkappa)$ in Appendix C shows that for average DIS on 
realistic nuclei $Q_{A}^2$ does not exceed several $(GeV/c)^2$, hence
this regime is a somewhat academic one, see however
the subsequent section 6. Still, let us assume that
$Q_A$ is so large that jets with $p_{\pm} \lsim Q_A$ are measurable.

Notice, that $|\bkappa_{i}|\sim Q_A$, so that one can neglect
$\bp_-$ in the photon's wave functions and the decorrelation momentum
$\bDelta$ in the
argument of $\Phi(\beta\nu_A(\bb),\bDelta -\bkappa_3 -\bkappa)$.
Then  an approximation 
\bea
&&\Biggl|\int d^2\bkappa_1\Phi((1-\beta)\nu_A(\bb),\bkappa_1)
\left\{\langle \gamma^*|z,\bp_{-} +\bkappa_1 +\bkappa_3\rangle -
\langle \gamma^*| z,\bp_{-} +\bkappa_1 +\bkappa_3+\bkappa\rangle\right\}\Biggr|^2
\nonumber \\
&&  
\approx \Bigl|\langle\gamma^*| z, \bkappa_3 \rangle
-
\langle \gamma^*|z, \bkappa_3+\bkappa \rangle\Bigr|^2 
\label{eq:5.5}
\eea
will be justified in (\ref{eq:5.4}).
The principal point is that the minijet-minijet inclusive cross
section depends
on neither the minijet nor decorrelation momentum, which proves
a disappearance of the azimuthal decorrelation
of minijets with the transverse momentum below the saturation scale.


\section{Azimuthal decorrelation of dijets in DIS off nuclei:
numerical estimates}

\begin{figure}[!t]
\begin{center}
   \epsfig{file=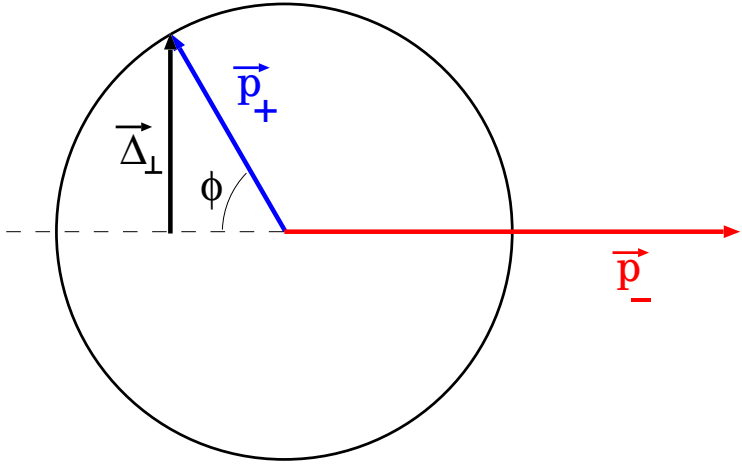,width=10cm}
\caption{\it The definition of the considered 
dijet configurations
and of the transverse component of the acoplanarity
momentum $\bDelta_{\perp}$.}
\end{center}
\end{figure}

The azimuthal decorrelation of two jets is quantified by the
mean transverse acoplanarity momentum squared 
$\langle \bDelta_{\perp}^2(\bb) \rangle$, where $\bDelta_{\perp}$ is
transverse to an axis of the jet with higher momentum, see fig.~3. 
Here we present numerical estimates for hard dijets, $|\bp_+| \gg Q_A$.
The
convolution property of the hard dijet cross section 
(\ref{eq:4.9}) suggests
\bea
\langle \bDelta_{\perp}^2(\bb) \rangle_{A} 
&=& \left\{ \int_{\cal C} d^2\bp_{-}\,\bDelta_{\perp}^2 {d\sigma_N \over dz d^2\bp_+d^2\bp_{-}}
\right\}
{\Biggr/}\left\{\int_{\cal C} d^2\bp_{-}{d\sigma_N \over dz d^2\bp_+d^2\bp_{-}}\right\}
\nonumber\\
&\approx&    
\langle \bkappa_{\perp}^2(\bb) \rangle_A  +  
\langle \bDelta_{\perp}^2 \rangle_N  \, ,
\label{eq:6.1}
\eea
where $\langle \bDelta_{\perp}^2 \rangle_N$ refers to DIS
on a free nucleon, 
  and $ \langle\bkappa_{\perp}^2(\bb) \rangle_A $ is the nuclear
broadening term 
\beq
\langle\bkappa_{\perp}^2(\bb) \rangle_A  =
\left\{\int_{\cal C} d^2\bkappa \, \bkappa_{\perp}^2
\Phi(2\beta \lambda_c  \nu_A(\bb),\bkappa)\right\} \Biggr/
\left\{\int_{\cal C} d^2\bkappa \, 
\Phi(2\beta \lambda_c  \nu_A(\bb),\bkappa)\right\} \, .
\label{eq:6.2}
\eeq
The sign $\approx$ in (\ref{eq:6.1}) 
reflects the kinematical limitations ${\cal C}$ on $\bp_-$ and 
$\bkappa$ in the practical 
evaluation of the acoplanarity distribution. In a typical final 
state shown in fig.~3  it is the harder jet with
larger transverse momentum which defines the jet axis and the
acoplanarity momentum $\bDelta$ will be defined in terms 
of components of the momentum of softer jet with respect to
that axis, for instance, see \cite{RHIC_STAR}. 
For the sake of definiteness, we present numerical 
estimates for the Gedanken experiment
in which we classify the event as a dijet if the quark 
and antiquark are produced in different hemispheres, i.e., if the
azimuthal angle $\phi$ between two jets exceeds $\pi/2$, the quark jet 
has fixed $|\bp_+|$ and  the antiquark jet has higher transverse momentum   
$|\bp_+| \lsim |\bp_-| \lsim 10 |\bp_+|$ 
(in the discussion of the experimental data
one often refers to the higher momentum jet as the trigger jet
and the softer jet as the away jet \cite{RHIC_STAR}).

\begin{figure}[!t]
\begin{center}
   \epsfig{file=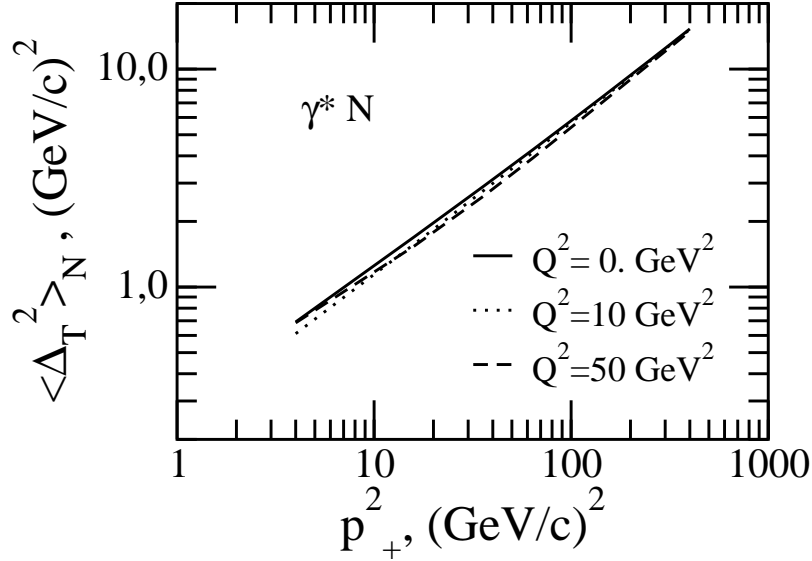,width=13cm,angle=270}
\caption{\it The mean acoplanarity momentum squared 
$\langle \bDelta _{\perp}^2\rangle_N$ for DIS on a free nucleon
target with production of trigger jets with the transverse
momentum higher than $\bp_{+}$ for several values of $Q^{2}$.
 The 
numerical results are for $x=0.01$ and the input
unintegrated gluon structure of the proton is taken from ref.
\protect{\cite{INDiffGlue}}.
}
\end{center}
\end{figure}

The free-nucleon quantity $\langle \bDelta_{\perp}^2 \rangle_N$ 
is evaluated from eq.~(\ref{eq:6.1}) with the free nucleon 
cross section (\ref{eq:2.7}).
For the evaluation purposes one can start with the  
small-$\bDelta$ expansion for excitation of 
hard, $\bp_+^2 \gg \varepsilon^2=z(1-z)Q^2$,  light flavor
dijets from transverse photons   
\bea
{d\sigma_N \over dz d^2\bp_+ d^2\bDelta} & \approx &
{1\over \pi}  e_f^2 \alpha_{em}\alpha_S(\bp_+^2)\left[z^2 + (1-z)^2\right]
\nonumber\\
& \times & {1\over \Delta^4} 
\cdot{\partial G(x,\bDelta^2)\over \partial \log \bDelta^2}\cdot 
{\bDelta^2 \over (\varepsilon^2 +\bp_+^2)(\varepsilon^2 +\bp_+^2+\bDelta^2 )} 
\, .
\label{eq:6.7}
\eea
The form of the last factor in
(\ref{eq:6.7})  only mimics its leveling off at $\bDelta^2 \gsim \bp_+^2$,
see eq.~ (\ref{eq:2.7}). Then in the denominator of (\ref{eq:6.1}) one 
finds the typical logarithmic integral
\beq
{1\over \pi} \int_{0}^{\pi} d\phi \int^{\bp_+^2} {d\bDelta^2 \over \bDelta^2} \cdot
{\partial G(x,\bDelta^2)\over \partial \log \bDelta^2} = G(x,\bp_+^2)
\label{eq:6.8}
\eeq
compared to the numerator of the form 
 \beq
{1\over \pi} \int_{0}^{\pi} \, d\phi \sin^2\phi \int^{\bp_+^2} d\bDelta^2 \cdot
{\partial G(x,\bDelta^2)\over \partial \log \bDelta^2} \sim {1\over 2} \bp_+^2 
{\cal F}(x,\bp_+^2)\, .
\label{eq:6.9}
\eeq
More accurate numerical estimates for
the selection criterions of our Gedanken experiment 
suggest the numerical factor $\approx 0.7$ in 
(\ref{eq:6.9}) and  
\bea
\langle \bDelta^2_{\perp} \rangle_N &=& \left\{
\int_{p_+} d^2\bp_{-}\,\bDelta_{\perp}^2 {d\sigma_N \over dz d^2\bp_+d^2\bp_{-}}
\right\}
{\Biggr/} \left\{\int_{p_+} d^2\bp_{-}{d\sigma_N \over dz d^2\bp_+d^2\bp_{-}}
\right\}
\nonumber\\
& \approx &{0.7}\cdot
{{\cal F}(x,\bp_+^2) \over G(x,\bp_+^2)}\bp_+^2 \, ,
\label{eq:6.10}
\eea
correctly describes the numerical results 
shown in fig.~4. As far as the dijets are hard, 
$\bp_{+}^2 \gsim z(1-z)Q^2 \sim {1\over 4}Q^2$, the acoplanarity
momentum distribution would not depend on $Q^2$, which 
holds still better if one considers $\sigma_T +\sigma_L$.
This point is illustrated in fig.~4, where we show 
$\langle \bDelta^2_{\perp} \rangle_N$ at $z=1/2$ for several values of
$Q^2$. Because of this weak dependence on $Q^2$ herebelow 
we make no distinction between DIS and real photoproduction, $Q^2=0$.


\begin{figure}[!t]
\begin{center}
   \epsfig{file=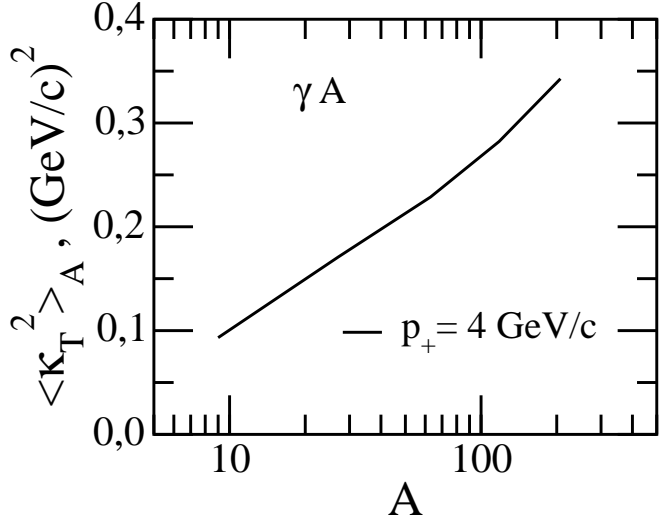,width=13cm, angle=270}
\caption{\it The atomic mass number
dependence of nuclear broadening contribution, 
$\langle \bkappa _{\perp}^2(\bb)\rangle_A$,
 to the
mean acoplanarity momentum squared for real
photoproduction
off nuclei  at $x=0.01$. The input
unintegrated gluon SF of the proton is taken from ref.
\protect{\cite{INDiffGlue}.}
}
\end{center}
\end{figure}

In the practical evaluations of the nuclear contribution
$\langle \bkappa_{\perp}^2(\bb) \rangle_A$ 
one can use an explicit expansion
\bea
\int_0^1 d \beta \Phi(2\beta \lambda_c  \nu_A(\bb),\bkappa) =
\sum_{j=0}^{\infty}w_A(\bb,j)f^{(j)}(\bkappa)=
\sum_{j=0}^{\infty} {1 \over j!} {\gamma(j+1,2\lambda_c  \nu_A(\bb)) \over
2\lambda_c  \nu_A(\bb)}
f^{(j)}(\bkappa)\, ,
\label{eq:6.3}
\eea
where $\gamma(j,x)= \int_{0}^x dy y^{j-1}\exp(-y)$ is an 
incomplete gamma-function.
The properties of the collective glue for $j$ overlapping
nucleons, $f^{(j)}(\bkappa)$,
are presented in Appendix C. For a heavy nucleus
(\ref{eq:6.3}) can be approximated by its integrand at
$\beta \approx 1/2$, ie., by $\Phi(\lambda_c  \nu_A(\bb),\bkappa)$. 
A bit more accurate evaluation of the numerically important
no-broadening contribution
from $j=0$ gives  
\bea
\int_0^1 d \beta \Phi(2\beta \lambda_c  \nu_A(\bb),\bkappa) \approx
w_A(\bb,0) \delta(\bkappa)+
(1-w_A(\bb,0)){1\over \pi} 
{\lambda_c Q_A^2(\bb) \over (\bkappa^2 +\lambda_c Q_A^2(\bb) )^2}\, 
\label{eq:6.4}
\eea
where $Q_A^2$ is given by eq.~(\ref{eq:C.12}) and
\beq
w_A(\bb,0) = {1 - \exp[-\nu_A(\bb)] \over \nu_A(\bb)}
\label{eq:6.4*}
\eeq
is a probability of the no-broadening contribution, which
for realistic nuclei is still substantial. 
In our Gedanken experiment $\langle \bkappa_{\perp}^2 (\bb)\rangle_A$  
must be evaluated over the constrained phase space ${\cal C}$, 
$\kappa_{\perp} \leq |\bp_+|$ and $\kappa_{L}>0$, and 
the analytic parameterization (\ref{eq:6.4}) gives  
\bea
\langle \bkappa_{\perp}^2 (\bb)\rangle_A 
&\approx
&\lambda_cQ_A^2(\bb)\cdot
 \left[ \log \tan\left( {\pi \over 4} + {1\over 2} \arctan {p_+ 
\over \sqrt{\lambda_c}Q_A(\bb)}
\right) - {p_+\over \sqrt{\lambda_cQ_A^2(\bb) + p_+^2}}\right]\nonumber\\
&\times& 
{ (1-w_{A}(\bb,0))\sqrt{\lambda_cQ_A^2(\bb) + p_+^2} 
\over w_A(\bb,0)\sqrt{\lambda_cQ_A^2(\bb) + p_+^2} + (1-w_A(\bb,0))p_+}
\, .
\label{eq:6.5} 
\eea 
We recall that (\ref{eq:6.1}) and (\ref{eq:6.5}) must only be used
for $|\bp_+| \gg Q_A(\bb)$.


\begin{figure}[!t]
\begin{center}
   \epsfig{file=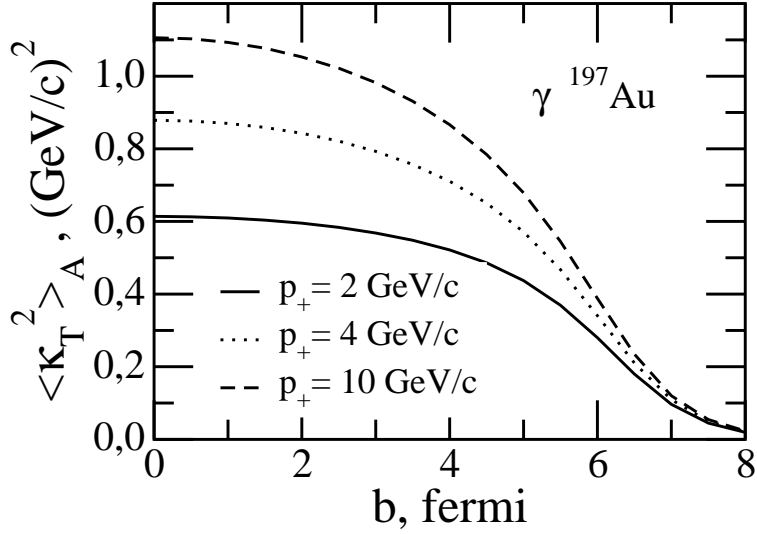,width=13cm, angle=270}
\caption{\it The impact parameter dependence of the 
nuclear broadening contribution, 
$\langle \bkappa _{\perp}^2(\bb)\rangle_A$,
 to the
mean acoplanarity momentum squared from 
peripheral DIS at large impact parameter 
to central DIS at $\bb=0$ for several
values of the away jet momentum $p_+$. The 
numerical results are for $x=0.01$ 
and the input
unintegrated gluon SF of the proton is taken from ref.
\protect{\cite{INDiffGlue}.}
}
\end{center}
\end{figure}

For average DIS on heavy nuclei the reference value is 
$\langle Q_{Au}^2(\bb)\rangle =0.9$ $(GeV/c)^2$,
see Appendix C. The atomic mass number dependence of 
nuclear broadening 
$\langle \bkappa_{\perp}^2 \rangle_A$ for jets with  
$p_+ = 4$ GeV/c in average DIS off nucleus
is shown in fig.~5. The principal reason why 
$\langle \bkappa_{\perp}^2 \rangle_A$ is numerically
small compared to $\langle Q_{Au}^2(\bb)\rangle$ is
that even for such a heavy nucleus as $^{197}Au$ 
the no-broadening probability in average DIS is large,
$\langle w_{Au}(\bb,0)\rangle \approx 0.5$.  
A comparison of the free nucleon broadening $\langle \bDelta_{\perp}^2 
\rangle_N$ from fig.~4 with the nuclear
contribution $\langle \bkappa_{\perp}^2 (\bb)
\rangle_A$ from fig.~5  shows suggest that 
the nuclear mass number dependence of azimuthal decorrelation of 
dijets in average DIS off nuclei will be relatively weak. 

However, nuclear broadening will be  substantially stronger
for a subsample of central DIS events at $\bb \sim 0$.
In fig.~6
we show for the gold, $^{197}Au$, target 
a dependence of the $\beta$ averaged nuclear broadening 
$\langle \bkappa_{\perp}^2 (\bb)\rangle_A$ on the
impact parameter at several
values of $p_+$. There are two related sources of the
$p_+$ dependence of $\langle \bkappa_{\perp}^2 (\bb)\rangle_A$.
First, since for hard dijets $r,r' \sim 1/p_+$, the 
strong coupling enters eq.~(\ref{eq:4.7}) and eq.~(\ref{eq:C.12})
as $\alpha_S(\bp_+^2)$. Then for hard jets 
$\nu_A(\bb) \propto \alpha_S(\bp_+^2)$ and $w_{A}(\bb=0,0)$ 
rises substantially with $p_+$ in the region of $p_+$
of the practical interest, $ 1 \lsim  p_+ \lsim $(5-10) GeV/c,
where the strong coupling varies rapidly. For a nucleus 
with mass number $A=200$
it rises
from $w_A(\bb=0,0)\approx 0.12$ at $ p_+ = $2 GeV/c to $\approx 0.20$ 
at 4 GeV/c to $\approx 0.25$ at $ p_+ = $10 GeV/c (for the nuclear
density parameterization see \cite{DeJaeger}).
Second, for the same reason that $\nu_A(\bb) \propto \alpha_S(\bp_+^2)$
the contribution from large $j$ in (\ref{eq:6.3}), and $ Q_A^2(\bb)$
thereof, diminish gradually with rising $p_+$, proportionally to
$\alpha_S(\bp_+^2)/\alpha_S(Q_A^2)$. In the region of $p_+ \lsim 10$ GeV/c 
of the practical interest one finds
$\langle \bkappa_{\perp}^2(\bb) \rangle_A \sim Q_A^2(\bb)$.

Now compare the numerical results in fig.~5 and fig.~6 for
$p_+=4$ GeV/c and $^{197}Au$ target. 
According to  eq.~(\ref{eq:C.13}) of Appendix C,  
\beq
Q_{A}^2(0) = ({4\over 3} -2) \langle Q_{A}^2(\bb) \rangle 
\, .
\label{eq:6.6}
\eeq
The no-broadening probability $w_{Au}(\bb,0)$
for central DIS is substantially smaller, 
$w_A(\bb=0,0)\approx 0.20$, than for average
DIS, $\langle w_{Au}(\bb,0)\rangle \approx 0.5$. 
In conjunction with (\ref{eq:6.6}) that entails 
an enhancement of $\langle \bkappa_{\perp}^2 (\bb)\rangle_A$  
by the factor of $\sim (2.5-3)$ from average to central DIS.
The same point is illustrated by the expectation value of $j$ in 
(\ref{eq:6.3}) for the $Au$ target, which for jets with 
$p_+=4$ GeV/c drops the 
the factor $\sim 3$ 
from $\langle j (\bb=0) \rangle =2.86$ to
$\langle j \rangle_A =0.87$ from central to average DIS.

One can enhance $Q_{A}^2$ and nuclear contribution
$\langle \bkappa_{\perp}^2 (\bb)\rangle_A$
still further selecting DIS events when
the photon breaks up into the $q\bar{q}$ pair on the front face 
of a nucleus, which in the language of (\ref{eq:4.10})
corresponds to the contribution from $\beta\to 1$, see the
discussion of (\ref{eq:6.3}).
Experimentally, precisely such events are isolated by
selecting very large multiplicity or very high transverse 
energy of produced secondary particles
(\cite{RHIC_STAR} and references therein).
Then, eq.~(\ref{eq:4.10}),  
shows, (see also a discussion of $\beta \approx 1/2$ approximation
in (\ref{eq:6.3})) 
that for very high multiplicity central DIS off Au nucleus 
$Q_A^2 \sim 2.5$ GeV$^2$ is quite feasible. Eq. ~(\ref{eq:6.5})
shows that for such a large $Q_A^2 \sim 2.5$ GeV$^2$
and $p_+=$(5-10) GeV  of the practical interest 
$\langle \bkappa_{\perp}^2(\bb=0) \rangle$ grows slower
than $\propto Q_A^2$, so that for high-multiplicity central 
DIS off $Au$ nucleus the value 
of $\langle \bkappa_{\perp}^2(\bb=0) \rangle$ will be  enhanced by the 
factor $\sim $(4-5) from 
$\langle \bkappa_{\perp}^2 \rangle_{Au}$ for
average DIS.

We have an overall good  understanding
of gross features of nuclear azimuthal decorrelations in
DIS off nuclei. Now we comment 
on the recent finding by the STAR collaboration 
of a disappearance of back-to-back high $p_{\perp}$ hadron 
correlation when going from peripheral to central
gold-gold collisions at RHIC \cite{RHIC_STAR}.
Our experience with application
of color dipole formalism to hard hadron-nucleus interactions
\cite{NPZcharm} suggests that our analysis of
acoplanarity of forward hard jets can be readily
generalized to mid-rapidity jets. One only has to choose an
appropriate system of dipoles, for instance, the open heavy
flavor production can be treated in terms of the
intranuclear propagation of the gluon-quark-antiquark system
in the overall color-singlet state. At RHIC energies jets
with moderately large $p_{\perp}$ are for the most part
due to gluon-gluon collisions. In our
language that can
be treated as a breakup of gluons into dijets and azimuthal
decorrelation of hard jets must be discussed in terms of
intranuclear propagation of color-octet gluon-gluon 
dipoles. For such gluon-gluon dipoles the relevant saturation scale
$Q_{8A}^2$ is larger \cite{NZ94} than that for the 
quark-antiquark dipoles
by the factor $2\lambda_c = C_A/C_F = 9/4$. Arguably, in
central nucleus-nucleus collisions distortions in the
target and projectile nuclei add up and the effective thickness
of nuclear matter is about twice that in DIS. 
Then,
the results shown in fig.~5 suggest that
for central gold-gold collisions the nuclear broadening 
of gluon-gluon dijets
could be quite substantial, 
$\langle \bkappa_{\perp}^2(\bb=0)\rangle_{AuAu} \sim $(3-4) $(GeV/c)^2$
for average central $AuAu$ collisions and even twice larger if
collisions take place at front surface of colliding nuclei. 
 
The 
principal effect of nuclear broadening is a reduction of the probability
of observing the back-to-back jets 
\beq 
\propto {\langle \bDelta_{\perp}^2
\rangle_N \over  
\langle \bkappa_{\perp}^2(\bb) 
\rangle_A + \langle \bDelta_{\perp}^2
\rangle_N}
\label{eq:6.11}
\eeq
and one needs to compare  
$\langle \bDelta_{\perp}^2 
\rangle_N$ to $\langle \bkappa_{\perp}^2 (\bb)
\rangle_A \, .
$
Our eq.~(\ref{eq:6.10}) for the free 
nucleon case holds as well for the gluon-gluon
collisions. Then the results shown in
fig.~3 entail that $\langle \bDelta_{\perp}^2 
\rangle_N \approx \langle \bkappa_{\perp}^2(0)\rangle_{AuAu} 
\sim $(3-4) $(GeV/c)^2$ at the jet momentum $p_+ = p_J =$ (6-8) GeV/c
and our nuclear
broadening 
would become substantial for all jets with $p_+$ below
the decorrelation
threshold momentum $p_J$.
In practice, the STAR  collaboration studied the azimuthal correlation of 
two high-$p_{\perp}$ hadrons and for the quantitative correspondence
between the STAR observable and azimuthal decorrelation in the parent 
dijet one needs to model fragmentation of jets into hadrons (for the 
modern fragmentation schemes see \cite{LUND}), here we notice that
the cutoff $p_+$ in our
Gedanken experiment is related to the momentum 
cutoff $p_{T,min}$ of  associated tracks from the away jet, 
whereas our jet of
momentum $\bp_-$ can be regarded as a counterpart of the trigger
jet of STAR. The STAR cutoff $p_T = $ 2GeV/c corresponds to the 
parents jets with the transverse momentum $p_+\sim (2-3)p_T
=$(4-6) GeV which is 
comparable to, or even smaller than, the decorrelation
threshold momentum
 $p_J= $ (6-8) GeV/c. Then eq. (\ref{eq:6.11}) suggests that 
in the kinematics of STAR the probability to observe the
back-to-back away and trigger jets decreases 
approximately twofold, and 
perhaps even stronger, form peripheral to central $Au Au$
collisions, so that  
our azimuthal decorrelation may contribute substantially
to the STAR effect. 

In practical consideration 
of azimuthal decorrelations
in central heavy ion collisions the above
distortions of the produced jet-jet
inclusive spectrum due to interactions with the nucleons of the
target and projectile ions must be complemented by
rescatterings of
parent high-$p_{\perp}$ partons on the abundantly produced
secondary hadrons. Our nuclear
decorrelation effect will be the dominant one 
and reinteractions with secondary  particles
will be marginal in $pA$ collisions, where for central
collisions we 
expect $\langle \bkappa_{\perp}^2(0)\rangle_{p Au} \sim 1.5$ $(GeV/c)^2$
and for central collisions in the regime of $\beta\to 1$, i.e.,
with limiting high multiplicity even 
$\langle \bkappa_{\perp}^2(0)\rangle_{p Au} \sim 3$ $(GeV/c)^2$ is
feasible.


\section{Nuclear $k_{\perp}$-factorization for 
$1/N_c^2$ corrections to the photon breakup}

Having established nuclear $k_{\perp}$-factorization properties
of  dijet cross section to leading order of the large-$N_c$ 
approximation, 
we turn to the $1/N_c^2$ corrections and  demonstrate
that with one simple 
exception the $1/N_c^2$ expansion can be regarded
as the higher-twist expansion.
The two sources of the $1/N_c^2$ corrections to the nuclear
distortion factor are higher order
terms in the off-diagonal $\sigma_{18}$ and the $\propto 1/(N_c^2-1)$
terms in $\sigma_{88}$, eq.~(\ref{eq:3.14}). 
To this end we notice that $\sigma_{88}$ can
be decomposed as
\bea
\sigma_{88}& =& \sigma(\bs)+
\sigma(\bs-\br+\br') + {2\Sigma_{18}(\bs,\br,\br')
\over N_c^2-1}\nonumber\\
&+&
 {\sigma(\bs)+
\sigma(\bs-\br+\br')-\sigma(\br)-\sigma(\br') \over N_c^2-1} \nonumber\\
&=&  {N_c^2  \over N_c^2-1}[\sigma(\bs)+
\sigma(\bs-\br+\br')] + {2\Sigma_{18}(\bs,\br,\br')
\over N_c^2-1}
-{\Delta \Sigma_{88}(\br,\br')
\over N_c^2-1} \, ,
\label{eq:7.1}
\eea
where
\bea
\Delta \Sigma_{88}(\br,\br') = 
\sigma(\br)+\sigma(\br')
\label{eq:7.2}
\eea
and we exactly reabsorbed one part of the $1/N_c^2$ correction 
into the leading large-$N_c$ term  of $\sigma_{88}$
scaling it up by the color factor $\lambda_c$. 

After some algebra one finds
\bea
&&\langle 11|S_{4A}(\bb_+',\bb_-',\bb_+,\bb_-)|11\rangle  =
\exp\left\{-{1\over 2}[\sigma(\br)+\sigma(\br')]T(\bb)\right\} \nonumber\\
&+& {\Sigma_{18}^2 (\bs,\br,\br')T^2(\bb) \over 4(N_c^2 -1)} 
\int_0^1 d\beta \int_{0}^{\beta} d\beta_1 \nonumber\\
&\times & \exp\left\{-{1\over 2}(1-\beta +\beta_1)
[\sigma(\br)+\sigma(\br')]T(\bb)\right\} \nonumber\\
&\times&\exp\left\{-{1\over 2}(\beta -\beta_1)
[\sigma(\bs)+\sigma(\bs -\br +\br')]T(\bb)\right\} \, .
\label{eq:7.3}
\eea
Notice, that the first term in (\ref{eq:7.3}) is cancelled by the
subtraction of the coherent diffractive term (\ref{eq:3.15}) in
(\ref{eq:3.1}), (\ref{eq:3.18}), so that only the subleading, $\propto
1/(N_c^2-1)$ term in (\ref{eq:7.3}) contributes to the dijet cross
section. The evaluation of corrections to the leading term
of the Sylvester expansion is a bit more complicated:  
\bea
&&\sqrt{N_c^2-1}\langle 88|S_{4A}(\bb_+',\bb_-',\bb_+,\bb_-)|11\rangle  = 
{1\over 2}\Sigma_{18} (\bs,\br,\br')T(\bb)\nonumber\\
&\times& \Biggl[
\int_0^1 d\beta 
\exp\left\{-{1\over 2}\beta 
[\sigma(\br)+\sigma(\br')]T(\bb)\right\} 
\exp\left\{-{1\over 2}(1-\beta) 
\sigma_{88}T(\bb)\right\} \nonumber\\
&+& {\Sigma_{18}^2 (\bs,\br,\br')T^2(\bb) \over 4(N_c^2 -1)} 
\int_0^1 d\beta \int_{0}^{\beta} d\beta_1 \int_{0}^{\beta_1} d\beta_2\nonumber\\
&\times&\exp\left\{-{1\over 2}(\beta -\beta_1+\beta_2)
[\sigma(\br)+\sigma(\br')]T(\bb)\right\} \nonumber\\
&\times&\exp\left\{-{1\over 2}(1-\beta +\beta_1-\beta_2)
[\sigma(\bs)+\sigma(\bs -\br +\br')]T(\bb)\right\} \Biggr]\, .
\label{eq:7.4}
\eea

The first term of (\ref{eq:7.4}) contains the attenuation factor
where $\sigma_{88}$ is still an exact diagonal matrix element, and
one must isolate the leading term and $1/(N_c^2-1)$ correction:
\bea
&&\exp\left\{-{1\over 2}(1-\beta) 
\sigma_{88}T(\bb)\right\} = \exp\left\{-{1\over 2}(1-\beta)\lambda_c
[\sigma(\bs)+\sigma(\bs -\br +\br')]T(\bb)\right\} \nonumber\\
&\times & \left\{
1 - {(1-\beta)\Sigma_{18} (\bs,\br,\br')T(\bb)\over N_c^2 -1 }
+ {(1-\beta)\Delta \Sigma_{88} (\br,\br')T(\bb) \over 2(N_c^2 -1)}
\right\}\, .
\label{eq:7.5}
\eea
The fundamental reason why the different components of
the second term $\propto 1/(N_c^2-1)$ in eq. (\ref{eq:3.14})
were treated differently is that the NSS representation with
positive valued Fourier transform only holds
for attenuating exponentials of the dipole cross section.
One can readily write down the related expansion for rising exponential 
$\exp[{1\over 2}\sigma(\br)T(\bb)]$, but its Fourier transform shall be
a sign-oscillating expansion:
\bea 
\exp\left[{1\over 2}\sigma(\bs)
T(\bb)\right] &=& \exp\left[2\nu_{A}(\bb)\right]
\sum_{j=0}^{\infty} (-1)^{j} w_{j}(\nu_{A}(\bb))\int d^2\bkappa
f^{(j)}(\bkappa)\exp(i\bkappa\bs)\, .
\label{eq:7.6}
\eea
For this reason combining together in the first term of
(\ref{eq:7.3}) the two exponentials with similar exponents 
$\propto [\sigma(\br)+\sigma(\br')]$ 
is not warranted,
because in the course of the $\beta$ integration
the sign of the exponent will change from attenuation to growth,
\bea
 \beta [\sigma(\br)+\sigma(\br')] -
{1\over N_c^2 -1}(1-\beta)\Delta\Sigma_{88}(\br,\br')
= {N_c^2\beta - 1 \over N_c^2 -1}[\sigma(\br)+\sigma(\br')]\, , 
\label{eq:7.7}
\eea
and it is advisable to stick to the perturbative expansion (\ref{eq:7.5}).

The final result for the nuclear absorption factor to an
accuracy $1/(N_c^2-1)$ reads
\bea
D_A(\bs,\br,\br',\bb) &=&{\Sigma_{18}(\bs,\br,\br')T^(\bb) \over 2(N_c^2 -1)} 
\int_0^1 d\beta \int_{0}^{\beta} d\beta_1 \nonumber\\
&\times&\exp\left\{-{1\over 2}(1-\beta +\beta_1)
[\sigma(\br)+\sigma(\br')]T(\bb)\right\} \nonumber\\
&\times&\exp\left\{-{1\over 2}(\beta -\beta_1)
[\sigma(\bs)+\sigma(\bs -\br +\br')]T(\bb)\right\} 
\label{eq:7.8}
\\
&+& {\Sigma_{18}^2 (\bs,\br,\br')T^2(\bb) \over 4(N_c^2 -1)} 
\int_0^1 d\beta \int_{0}^{\beta} d\beta_1 \int_{0}^{\beta_1} d\beta_2\nonumber\\
&\times&\exp\left\{-{1\over 2}(\beta -\beta_1+\beta_2)
[\sigma(\br)+\sigma(\br')]T(\bb)\right\} \nonumber\\
&\times&\exp\left\{-{1\over 2}(1-\beta +\beta_1-\beta_2)
[\sigma(\bs)+\sigma(\bs -\br +\br')]T(\bb)\right\} \Bigr\}
\label{eq:7.9}\\
&-&{\Sigma_{18} (\bs,\br,\br')T(\bb)\over N_c^2 -1 }
\int_0^1 d\beta (1-\beta)\nonumber\\
&\times&\exp\left\{-{1\over 2}\beta 
[\sigma(\br)+\sigma(\br')]T(\bb)\right\} \nonumber\\
&\times&\exp\left\{-{1\over 2}(1 -\beta)\lambda_c
[\sigma(\bs)+\sigma(\bs -\br +\br')]T(\bb)\right\}
\label{eq:7.10}
\\
&+& {\Delta\Sigma_{88}(\br,\br') T(\bb) \over 2(N_c^2 -1)}
\int_0^1 d\beta (1-\beta) \nonumber\\
&\times& \exp\left\{-{1\over 2}\beta 
[\sigma(\br)+\sigma(\br')]T(\bb)\right\}\nonumber\\
&\times&\exp\left\{-{1\over 2}(1 -\beta)\lambda_c
[\sigma(\bs)+\sigma(\bs -\br +\br')]T(\bb)\right\}
\label{eq:7.11}
\\
&+&
\int_0^1 d \beta \exp\left\{-{1\over 2}(1-\beta)[\sigma(\br)+\sigma(\br')]
T(\bb)\right\}  \nonumber\\
&\times&\exp\left\{-{1\over 2}\beta[\sigma(\bs)+\sigma(\bs+\br'-\br)]
T(\bb)\right\} \, .
\label{eq:7.12}
\eea
Here the last term (\ref{eq:7.12})
is the leading large-$N_c$ result, the first (\ref{eq:7.8})
and second (\ref{eq:7.9}) terms describe contributions to the dijet cross section
of  second and third order in the off-diagonal 
$\sigma_{18}$, the third (\ref{eq:7.10}) and fourth (\ref{eq:7.11}) term 
come from the expansion (\ref{eq:7.5}).

As in illustration of salient features of $1/(N_c^2-1)$ corrections
we expose in details the contribution from the first term
(\ref{eq:7.8}). 
Following the considerations in sections 4 \& 5, one readily
obtains
\bea
&&{d\Delta \sigma_{in}^{(1)} 
\over d^2\bb dz d\bp_{-} d\bDelta} = {\alpha_S^2 \sigma_0^2
T^2 (\bb)\over 4(2\pi)^2 (N_c^2 -1)}\int_0^1 d \beta \int_0^{\beta}d\beta_1
\nonumber \\
&\times& \int d^2\bq_1 d^2\bq_2  d^2\bkappa_3
f(\bq_1)f(\bq_2)\nonumber\\
&\times&
\Phi((\beta-\beta_1)\nu_A(\bb),\bDelta -\bkappa_3 -\bq_1-\bq_2)
\Phi((\beta-\beta_1)\nu_A(\bb),\bkappa_3)
\nonumber\\
&\times &
\Biggl|\int d^2\bkappa_1 \Phi((1-\beta+\beta_1)\nu_A(\bb),\bkappa_1)\nonumber\\
&\times &
\Bigl\{\langle \gamma^*|z,\bp_{-} +\bkappa_1 +\bkappa_3\rangle \nonumber
-
\langle \gamma^*|z,\bp_{-} +\bkappa_1 +\bkappa_3+\bq_1\rangle \nonumber\\
&-&\langle \gamma^*|z,\bp_{-} +\bkappa_1 +\bkappa_3 +\bq_2\rangle +
\langle \gamma^*|z,\bp_{-} +\bkappa_1 +\bkappa_3+\bq_1 +\bq_2\rangle\Bigr\}\Biggr|^2\, .
\label{eq:7.13}
\eea
Of particular interest is the large-$|\bp_-|$ behavior of (\ref{eq:7.13}).
We notice that for $\bp_-^2 \gg Q_A^2(\bb)$ one can neglect $\bkappa_{1,3}$ in
the argument of the photon wave function, so that
\bea
&&\int d^2\bkappa_1 \Phi((1-\beta+\beta_1)\nu_A(\bb),\bkappa_1)\nonumber\\
&\times&
\Bigl\{\langle \gamma^*|z,\bp_{-} +\bkappa_1 +\bkappa_3\rangle -
\langle \gamma^*|z,\bp_{-} +\bkappa_1 +\bkappa_3+\bq_1\rangle \nonumber\\
&-&\langle \gamma^*|z,\bp_{-} +\bkappa_1 +\bkappa_3 +\bq_2\rangle +
\langle \gamma^*|z,\bp_{-} +\bkappa_1 +\bkappa_3+\bq_1 +\bq_2\rangle\Bigr\}\nonumber\\
&\approx&
\Bigl\{\langle \gamma^*|z,\bp_{-}\rangle -
\langle \gamma^*|z,\bp_{-}+\bq_1\rangle 
-\langle \gamma^*|z,\bp_{-} +\bq_2\rangle +
\langle \gamma^*|z,\bp_{-} +\bq_1 +\bq_2\rangle\Bigr\}\,,
\label{eq:7.14}
\eea
where we used the normalization property
$\int d^2\bkappa_1 \Phi((1-\beta+\beta_1)\nu_A(\bb),\bkappa_1)=1$. 
Next one can readily verify that 
\bea
&&\int  d^2\bkappa_3
\Phi((\beta-\beta_1)\nu_A(\bb),\bDelta -\bkappa_3 -\bq_1-\bq_2)
\Phi((\beta-\beta_1)\nu_A(\bb),\bkappa_3)\nonumber\\
& =&
\Phi((\beta-\beta_1)\nu_A(\bb),\bDelta -\bq_1-\bq_2)
\label{eq:7.15}
\eea

Incidentally, by a similar analysis of the onset of high-$\bp_+$
limit one would obtain the linear nuclear $k_{\perp}$-factorization
(\ref{eq:4.10}) for hard dijets from the nonlinear
nuclear $k_{\perp}$-factorization (\ref{eq:5.4}).

The combination of the photon wave functions in (\ref{eq:7.14})
corresponds to the second finite difference in $\bq_1,\bq_2$, so 
that for jets with $\bp_-^2 \gg \varepsilon^2$ one has an estimate
\bea
&&\Bigl|\langle \gamma^*|z,\bp_{-}\rangle -
\langle \gamma^*|z,\bp_{-}+\bq_1\rangle 
+\langle \gamma^*|z,\bp_{-} +\bq_2\rangle -
\langle \gamma^*|z,\bp_{-} +\bq_1 +\bq_2\rangle\Bigr|^2\nonumber\\
&\approx& \Bigl|\langle \gamma^*|z,\bp_{-}\rangle\Bigr|^2 
\cdot {\bq_1^2\bq_2^2\over (\bp_-^2)^2}
\label{eq:7.16}
\eea
which shows that the contribution to the dijet cross section
from terms of second order in $\sigma_{18}^2$ is the higher twist
correction. Compared to the leading large-$N_c$ cross section,
it contains extra $\int d^2\bq_2\, \bq_2^2 f(\bq_2)$ and extra power
of $\alpha_S \sigma_{0} T(\bb)$ which combine  to precisely
the dimensional nuclear
saturation scale $Q_{A}^2(\bb)$, see eq.~(\ref{eq:6.5}), so that
the resulting suppression factor is
\beq
{d\Delta \sigma_{in}^{(1)}\over 
d\sigma_{in}} \sim {1 \over (N_c^2-1)} \cdot { Q_A^2(\bb) \over \bp_-^2}\, .
\label{eq:7.17}
\eeq

As far as an expansion in higher inverse powers of the hard
scale $\bp_-^2$ is concerned, $\Delta \sigma_{in}^{(1)}$ has a form 
of higher twist correction. 
In the retrospect, one observes that the principal 
approximation (\ref{eq:7.14}) in the above derivation for 
hard dijets amounts to putting $|\br|,|\br'| \ll |\bs|$
in the attenuation factors in the $\beta,\beta_1$ integrand
in (\ref{eq:7.8}). However, the exact $\br,\br'$ dependence
must be retained in the prefactor $\Sigma_{18} (\bs,\br,\br')$,
because it vanishes if either $\br=0$ or $\br'=0$. It is 
precisely the 
latter property which provides the finite difference 
structure of the combination of the photon wave functions
in (\ref{eq:7.13}), (\ref{eq:7.14}) and is behind the
higher twist property (\ref{eq:7.17})
of the $1/(N_c^2-1)$ correction.

The second term (\ref{eq:7.9}) gives the correction
\bea
&&{d\Delta \sigma_{in}^{(2)} 
\over d^2\bb dz d\bp_{-} d\bDelta} = {\alpha_S^3 \sigma_0^3
T^3 (\bb)\over 8(2\pi)^2 (N_c^2 -1)}\int_0^1 d \beta \int_0^{\beta}d\beta_1
\int_0^{\beta_1}d\beta_2
\nonumber \\
&\times& \int d^2\bq_1 d^2\bq_2  d^2\bq_3 d^2\bkappa_3
f(\bq_1)f(\bq_2)f(\bq_3)\nonumber\\
&\times&
\Phi((1-\beta+\beta_1-\beta_2)\nu_A(\bb),\bDelta -\bkappa_3 -\bq_1-\bq_2-\bq_3)
\Phi((1-\beta+\beta_1-\beta_2)\nu_A(\bb),\bkappa_3)
\nonumber\\
&\times &
\Biggl|\int d^2\bkappa_1 
\Phi((\beta-\beta_1+\beta_2)\nu_A(\bb),\bkappa_1)\nonumber\\
&\times &
\Bigl\{\langle \gamma^*|z,\bp_{-} +\bkappa_1 +\bkappa_3\rangle \nonumber
-
\langle \gamma^*|z,\bp_{-} +\bkappa_1 +\bkappa_3+\bq_1\rangle 
\nonumber\\
&-&\langle \gamma^*|z,\bp_{-} +\bkappa_1 +\bkappa_3 +\bq_2\rangle +
\langle \gamma^*|z,\bp_{-} +\bkappa_1 +\bkappa_3+\bq_1 +\bq_2\rangle
\nonumber\\
&-&\langle \gamma^*|z,\bp_{-} +\bkappa_1 +\bkappa_3 +\bq_3\rangle +
\langle \gamma^*|z,\bp_{-} +\bkappa_1 +\bkappa_3+\bq_3 +\bq_1\rangle
\nonumber\\
&+&\langle \gamma^*|z,\bp_{-} +\bkappa_1 +\bkappa_3 +\bq_3+\bq_2\rangle -
\langle \gamma^*|z,\bp_{-} +\bkappa_1 +\bkappa_3+\bq_1 +\bq_2+\bq_3|
\gamma^*\rangle\Bigr\}\Biggr|^2\, .
\label{eq:7.18}
\eea
The combination of the photon wave functions in (\ref{eq:7.18}) 
corresponds to the third finite derivative in $\bq_{1,2,3}$.
Starting from (\ref{eq:7.18}) one can readily repeat the 
analysis which lead to an estimate (\ref{eq:7.17}). 
Alternatively, one can take the simplified form of the attenuation
factors, as explained below eq.~(\ref{eq:7.17}). Either way we
find that contribution from third order terms in $\sigma_{18}$ is
of still higher twist and has
a smallness
\beq
{d\Delta \sigma_{in}^{(2)}\over 
d\sigma_{in}} \sim {1 \over (N_c^2-1)} \left({ Q_A^2(\bb) \over \bp_-^2}\right)^2\, .
\label{eq:7.19}
\eeq

Apart from the slight difference in the structure of the $\beta$ 
integrations, the correction (\ref{eq:7.10}) is not any different
from $d\Delta \sigma^{(1)}$ of eq. (\ref{eq:7.14}):
\bea
&&{d\Delta \sigma_{in}^{(3)} 
\over d^2\bb dz d\bp_{-} d\bDelta} = -{\alpha_S^2 \sigma_0^2
T^2 (\bb)\over 4(2\pi)^2 (N_c^2 -1)}\int_0^1 d \beta (1-\beta)
\nonumber \\
&\times& \int d^2\bq_1 d^2\bq_2  d^2\bkappa_3
f(\bq_1)f(\bq_2)\nonumber\\
&\times&
\Phi((1-\beta)\lambda_c \nu_A(\bb),\bDelta -\bkappa_3 -\bq_1-\bq_2)
\Phi((1-\beta)\lambda_c \nu_A(\bb),\bkappa_3)
\nonumber\\
&\times &
\Biggl|\int d^2\bkappa_1 \Phi(\beta)\nu_A(\bb),\bkappa_1)\nonumber\\
&\times &
\Bigl\{\langle \gamma^*|z,\bp_{-} +\bkappa_1 +\bkappa_3\rangle \nonumber
-
\langle \gamma^*|z,\bp_{-} +\bkappa_1 +\bkappa_3+\bq_1\rangle \nonumber\\
&-&\langle \gamma^*|z,\bp_{-} +\bkappa_1 +\bkappa_3 +\bq_2\rangle +
\langle \gamma^*|z,\bp_{-} +\bkappa_1 +\bkappa_3+\bq_1 +\bq_2\rangle\Bigr\}\Biggr|^2\, .
\label{eq:7.20}
\eea
Consequently 
the same estimate (\ref{eq:7.17}) holds also for $d\Delta \sigma^{(3)}$.

The correction $d\Delta \sigma^{(4)}$ needs a bit more scrutiny. 
It contains a product of the first and second finite derivatives 
of the photon wave function,
\bea
&&{d\Delta \sigma_{in}^{(4)} 
\over d^2\bb dz d\bp_{-} d\bDelta} = {\alpha_S^2 \sigma_0^2
T^2 (\bb)\over 2(2\pi)^2 (N_c^2 -1)}\int_0^1 d \beta (1-\beta)
\nonumber \\
&\times& \int d^2\bq_1 d^2\bq_2  d^2\bkappa_1  d^2\bkappa_2 d^2\bkappa_3
f(\bq_1)f(\bq_2)\Phi(\beta)\nu_A(\bb),\bkappa_1)\Phi(\beta)\nu_A(\bb),\bkappa_2)
\nonumber\\
&\times&
\Phi((1-\beta)\lambda_c \nu_A(\bb),\bDelta -\bkappa_3 -\bq_1-\bq_2)
\Phi((1-\beta)\lambda_c \nu_A(\bb),\bkappa_3)
\nonumber\\
&\times & 
\Bigl\{\langle \gamma^*|z,\gamma^*|z,\bp_{-} +\bkappa_1 +\bkappa_3\rangle 
-
\langle \gamma^*|z,\gamma^*|z,\bp_{-} +\bkappa_1 +\bkappa_3+\bq_1\rangle\Bigr\} \nonumber\\
&\times & 
\Bigl\{\langle \gamma^*|z,\bp_{-} +\bkappa_1 +\bkappa_3\rangle 
-
\langle \gamma^*|z,\bp_{-} +\bkappa_1 +\bkappa_3+\bq_1\rangle \nonumber\\
&-&\langle \gamma^*|z,\bp_{-} +\bkappa_1 +\bkappa_3 +\bq_2\rangle +
\langle \gamma^*|z,\bp_{-} +\bkappa_1 +\bkappa_3+\bq_1 +\bq_2\rangle\Bigr\}\, ,
\label{eq:7.21}
\eea
and in the interesting case of hard dijets 
\bea
&&{d\Delta \sigma_{in}^{(4)} 
\over d^2\bb dz d\bp_{-} d\bDelta} = {\alpha_S^2 \sigma_0^2
T^2 (\bb)\over 2(2\pi)^2 (N_c^2 -1)}\int_0^1 d \beta (1-\beta)
\nonumber \\
&\times& \int d^2\bq_1 d^2\bq_2  
f(\bq_1)f(\bq_2)
\Phi(2(1-\beta)\lambda_c \nu_A(\bb),\bDelta -\bq_1-\bq_2)
\nonumber\\
&\times & 
\Bigl\{\langle \gamma^*|z,\gamma^*|z,\bp_{-}\rangle 
-
\langle \gamma^*|z,\gamma^*|z,\bp_{-}+\bq_1\rangle\Bigr\} \nonumber\\
&\times & 
\Bigl\{\langle \gamma^*|z,\bp_{-}\rangle 
-
\langle \gamma^*|z,\bp_{-} +\bq_1\rangle 
-\langle \gamma^*|z,\bp_{-} +\bq_2\rangle +
\langle \gamma^*|z,\bp_{-} +\bq_1 +\bq_2\rangle\Bigr\}\, .
\label{eq:7.22}
\eea
The leading term of the small-$\bq_{1,2}$ expansion of the 
product of the 
photon wave functions in (\ref{eq:7.20}) is a 
quadratic function
of $\bq_1$ and  linear function of $\bq_2$ of the form
\beq
|\langle \gamma^*|z,\bp_{-}\rangle|^2 
{(\bp_- \bq_1)^2 (\bp_- \bq_2) \over p_-^6}\, .
\label{eq:7.23}
\eeq
The leading nonvanishing term comes from an expansion of
the nuclear WW glue
\bea
\Phi(2(1-\beta)\lambda_c \nu_A(\bb),\bDelta -\bq_1-\bq_2) -
\Phi(2(1-\beta)\lambda_c \nu_A(\bb),\bDelta) \nonumber\\
\sim \Phi(2(1-\beta)\lambda_c \nu_A(\bb),\bDelta) {(\bDelta ,\bq_1+\bq_2)
\over (2(1-\beta)\lambda_c Q_A^2(\bb) +\bDelta^2)}\,.
\label{eq:7.24}
\eea
Namely, upon the azimuthal averaging 
of (\ref{eq:7.23}) in conjunction with (\ref{eq:7.24})
we find the leading nonvanishing term of
the form $2(\bp_- \bq_2)(\bDelta \bq_2) 
\Longrightarrow (\bp_- \bDelta)\bq_2^2$, so that 
\beq
{d\Delta \sigma_{in}^{(4)}\over 
d\sigma_{in}} \sim {1 \over (N_c^2-1)} {(\bp_- \bDelta) \over \bp_-^2}\, ,
\label{eq:7.25}
\eeq
which is reminiscent of higher twist-3 correction.

To summarize, nonlinear nuclear $k_{\perp}$-factorization
allows a consistent evaluation of $1/N_c^2$ corrections.
We demonstrated how the expansion in $1/(N_c^2 -1)$ 
comes along with a higher twist expansion. One exception is
the reabsorption of one of the terms $\propto 1/(N_c^2 -1)$  
in $\sigma_{88}$ into the renormalization of the leading
term in  $\sigma_{88}$ by the $N_{c}$ dependent factor $\lambda_c$.
We conclude this discussion by a comment that all the arguments of
section 5 on disappearance of azimuthal correlations of minijets 
hold for the $1/N_c^2$ corrections as well.

\section*{Summary and conclusions}

We formulated the theory of the breakup of photons into dijets
in DIS off nuclear targets based on the
consistent treatment of propagation of color dipoles in nuclear
medium. The non-Abelian intranuclear evolution of color
dipoles gives rise to a nontrivial spectrum of the attenuation
eigenvalues, still the familiar Glauber-Gribov multiple-scattering
results  are recovered for the nuclear total
cross sections. However, for the more special cases like 
 inelastic DIS in which the photon breaks up into the
color singlet dijets, the cross section depends on the complete
spectrum of the attenuation eigenstates.

We derived the nuclear broadening of acoplanarity
momentum distribution
in the breakup of photons into dijets,
see eqs. (\ref{eq:4.9}), (\ref{eq:5.4}).
Our principal finding is that
all nuclear DIS observables - the amplitude of coherent
diffractive breakup into dijets \cite{NSSdijet}, nuclear sea 
quark SF
and its decomposition into equally important genuine inelastic
and diffractive components as carried out in \cite{Saturation}
and the jet-jet inclusive cross section derived in the present
paper, - are uniquely calculable in terms of the NSS-defined
collective nuclear WW glue. This
property can be regarded as a nuclear $k_{\perp}$-factorization theorem
which connects
DIS in the regimes of low and high density of partons.
For the generic dijet cross section nuclear $k_{\perp}$-factorization is
of highly nonlinear form which must be contrasted to the
linear hard factorization for the free nucleon target.
This result is derived to the leading order in large-$N_c$, the further 
evaluation of the $1/N_c^2$ corrections shows a close 
relation between the $1/N_c^2$ and high-twist expansions. 
Furthermore, the $1/N_c^2$ corrections do themselves 
admit the nonlinear nuclear $k_{\perp}$-factorization 
representation.  

We demonstrated a disappearance of
azimuthal jet-jet correlations of minijets with momenta below
the saturation scale. Based on the ideas \cite{NPZcharm,LPM} 
on generalization of the
dipole picture to hadron-nucleus collisions
we presented qualitative estimates of the broadening effect
for mid-rapidity jets produced in central nucleus-nucleus 
collisions and argued that our azimuthal decorrelation
may contribute substantially to a disappearance of
back-to-back high $p_{\perp}$ hadron correlation in central
gold-gold collisions as observed by the STAR collaboration 
at RHIC \cite{RHIC_STAR}. 

We conclude by the comment that all the results for
hard single-jet and jet-jet inclusive cross sections can be
readily extended from DIS to the breakup of projectile hadrons 
into forward jets. Indeed, as argued in \cite{NSSdijet},
the final state interaction between the final state quark and
antiquark can be neglected and plane-wave approximation becomes
applicable as soon as the invariant mass of the forward
jet system exceeds a
typical mass scale of prominent meson and baryon resonances.
The results will be presented elsewhere \cite{Pions}, here
we confine ourselves to the statement that 
although our principal point about nonlinear nuclear
$k_{\perp}$ factorization is fully retained, we find
important distinctions between the breakup of pointlike
photons and non-pointlike hadrons  

This work has been partly supported by the INTAS grants 97-30494
\& 00-00366 and the DFG grant 436RUS17/119/02
\pagebreak\\

\appendix

\section*{Appendix A: Calculation of the 4-body color dipole cross
section}

The Feynman diagrams for the matrix of 4-parton dipole cross section
$\sigma_4(\bs,\br,\br')$, eqs. (\ref{eq:3.12})-(\ref{eq:3.14}) 
are shown in fig.~2. The profile
function for the color-singlet $q\bar{q}$ pair is given by the
diagrams of fig. 2a-2c:
\bea
&&2\Gamma(fig. 2a-2c;( q\bar{q})_1 N; \bb_+,\bb_-)=
{1\over N_c}\delta_{ab}
\left\{[\chi^2(\bb_+)+\chi^2(\bb_-)] {\rm Tr}(T^aT^b)\right.\nonumber\\
&&-\left. 2\chi(\bb_+)\chi(\bb_-) {\rm Tr}(T^aT^b)\right\}
= {N_c^2 -1 \over 2N_c}[\chi(\bb_+)-\chi(\bb_-)]^2\,,
\label{eq:A.1}
\eea
which has already been cited in the main text, eq.~(\ref{eq:3.4}).
Upon adding the contribution from diagrams of fig.~2e-2h, we obtain
an obvious result
(\ref{eq:3.12}).

The color-diagonal contribution of the same diagrams to the interaction
of the color-octet $q\bar{q}$ pair with the nucleon equals
\bea
&&2\Gamma(fig. 2a-2c;( q\bar{q})_8 N; \bb_+,\bb_-)= \nonumber\\
&&{2\over N_c^2-1 }\delta_{ab}
\left\{
[\chi^2(\bb_+)+\chi^2(\bb_-)] {\rm Tr}(T^cT^aT^bT^c)
-2\chi(\bb_+)\chi(\bb_-) {\rm Tr}(T^cT^aT^cT^b)\right\}
\nonumber\\
&&= {N_c^2 -1 \over 2N_c}\left\{\left[\chi^2(\bb_+)+\chi^2(\bb_-)\right]
+ {2 \over N_c^2-1}\chi(\bb_+)\chi(\bb_-)\right\}\, .
\label{eq:A.2}
\eea
A contribution to the matrix element
$\langle 88|\sigma_4|88\rangle$ from color-diagonal interactions of the
$(q'\bar{q}')$ pair is obtained from (\ref{eq:A.3}) by the substitution
$\bb_{\pm} \to \bb_{\pm}'$:
\bea
&&\Gamma_4(fig. 2a-2c + fig. 2e-2h; (88)N; \bb_+,\bb_-, \bb_+',\bb_-')= \nonumber\\
&&\Gamma(fig 2a-2c; q\bar{q})_8 N;
\bb_+,\bb_-)+\Gamma(fig. 2a-2c; q\bar{q})_8 N; \bb_+',\bb_-')\, .
\label{eq:A.3}
\eea
The diagrams of fig. 2e-2h describe processes with  color-space rotation of the
$(q\bar{q})$ pair:
\bea
&&2\Gamma_{4}(fig.2i-2l; (88)N \to (88)N; \bb_+,\bb_-, \bb_+',\bb_-')= \nonumber\\
&&{8 \over N_c^2 -1}\delta_{ab} \left\{\left[\chi(\bb_+)\chi(\bb_-')
+ \chi(\bb_-)\chi(\bb_+')\right]
{\rm  Tr}(T^cT^aT^d){\rm  Tr}(T^cT^bT^d)\right.\nonumber \\
&&\left. -\left[\chi(\bb_+)\chi(\bb_+') +\chi(\bb_-)\chi(\bb_-')\right]
{\rm  Tr}(T^cT^aT^d){\rm  Tr}(T^dT^bT^c)\right\}\nonumber\\
&&= - {N_c^2 -1  \over N_c}
\left\{{2 \over N_c^2-1}\left[\chi(\bb_+)\chi(\bb_-')
+ \chi(\bb_-)\chi(\bb_+')\right] \right.\nonumber\\
&&\left. + {N_c^2 -2 \over N_c^2-1} \left[\chi(\bb_+)\chi(\bb_+') +
\chi(\bb_-)\chi(\bb_-')\right]\right\}\, .
\label{eq:A.4}
\eea

The $(11)N \to (88)N$ transition matrix element comes from diagrams of
fig. 2e-h:
\bea
&&2\Gamma_4(2i-l ; (11)N \to (88)N; \bb_+,\bb_-, \bb_+',\bb_-')= \nonumber\\
&&{4 \over N_c\sqrt{N_c^2 -1}}\delta_{ab} \left\{\left[\chi(\bb_+)\chi(\bb_-')
+ \chi(\bb_-)\chi(\bb_+')\right]
{\rm  Tr}(T^cT^a){\rm  Tr}(T^cT^b)\right.\nonumber \\
&&\left. -\left[\chi(\bb_+)\chi(\bb_+') +\chi(\bb_-)\chi(\bb_-')\right]
{\rm  Tr}(T^cT^a){\rm  Tr}(T^cT^b)\right\}\nonumber\\
&&= {N_c^2 -1  \over N_c} \cdot {1\over \sqrt{N_c^2 -1}}
 \left\{ \left[\chi(\bb_+)\chi(\bb_-')
+ \chi(\bb_-)\chi(\bb_+')\right]\right.\nonumber\\
&&\left.  -\left[\chi(\bb_+)\chi(\bb_+') +\chi(\bb_-)\chi(\bb_-')\right]\right\}
\, .
\label{eq:A.5}
\eea
Upon the rearrangement $-2\chi(\bb_i)\chi(\bb_j) =
[\chi(\bb_i)-\chi(\bb_j)]^2 -\chi^2(\bb_i)-\chi^2(\bb_j)$
one can readily verify that the terms $\propto \chi^2(\bb_i)$ 
cancel each other, and the 4-body cross section will be a linear combination
of $\sigma(\bb_i - \bb_j)$, recall a discussion in \cite{NZ94}.


\section*{Appendix B: Non-Abelian vs. Abelian 
aspects of intranuclear propagation of color dipoles
and Glauber-Gribov formalism}

The intranuclear propagation of color-octet $q\bar{q}$ pairs
is part and parcel of a complete formalism for DIS off nucleus. 
It is interesting to see how one recovers the quasi-Abelian
color-dipole results for the nuclear cross sections \cite{NZ91,NZZdiffr}
which are of the Glauber-Gribov form \cite{Glauber,Gribov}.
We consider first the total inelastic cross section obtained from
(\ref{eq:3.1}) upon the integration over the transverse momenta
$\bp_{\pm}$  of the quark and antiquark, which amounts
to putting $\bb_+=\bb_+'$ and $\bb_-=\bb_-'$. Then we are left with
the system of two color dipoles of the same size $\br=\bb_+ -\bb_- =
\br'=\bb_+' -\bb_-'$,
and the matrix of the 4-body cross section has the eigenvalues
\bea
\Sigma_1 = 0 \, ,
\label{eq:B.1}
\eea
\bea
\Sigma_2 = {2N_c^2 \over N_c^2 -1} \sigma(\br)
\label{eq:B.2}
\eea
with the eigenstates
\bea
|f_1\rangle = {1\over N_c}(|11\rangle + \sqrt{N_c^2-1} |88\rangle)\, ,
\label{eq:B.3}
\eea
\bea
|f_2\rangle = {1\over N_c}(\sqrt{N_c^2-1}|11\rangle - |88\rangle)\, .
\label{eq:B.4}
\eea
The
existence of the non-attenuating 4-quark state with $\Sigma_1=0$
is quite obvious and corresponds to an overlap of two $q\bar{q}$
dipoles
of the same size with neutralization of color charges. An
existence of such a non-attenuating state is shared by an
Abelian and non-Abelian quark-gluon interaction.
The intranuclear attenuation eigen-cross section (\ref{eq:B.2})
differs from $\sigma(\br)$, 
for the color-singlet $q\bar{q}$ pair by the nontrivial
color factor $2\lambda_c= 2N_c^2/(N_c^2-1)=C_{A}/C_{F}$ which
derives from the relevant 
4-parton state
being in the color octet-(anti)octet configuration.

The crucial point is that the final state which enters the calculation of
the genuine inelastic DIS off a nucleus, see eq. (\ref{eq:3.8}),
is precisely the eigenstate $|f_1\rangle$. Then, even without
invoking the Sylvester expansion (\ref{eq:3.17}), (\ref{eq:3.18}),
the straightforward result for  the inelastic cross section is
\bea 
\sigma_{in} &=& \int d^2\br dz |\Psi(Q^2,z,\br)|^2 \int d^2\bb
\left\{ N_c\langle f_1| \exp\left[-{1\over 2}\sigma_4
T(\bb)\right] |11\rangle
-\exp\left[-\sigma(\br) T(\bb)\right]\right\} \nonumber\\
&=& \int d^2\bb\langle \gamma^*| \left\{\exp\left[-{1\over
2}\Sigma_1 T(\bb)\right]
-\exp\left[-\sigma(\br) T(\bb)\right]\right\} |\gamma^* \rangle \nonumber\\
&=& \int d^2\bb\langle \gamma^*| \left\{1 -\exp\left[-\sigma(\br)
T(\bb)\right]\right\} |\gamma^*\rangle \label{eq:B.5} 
\eea 
what is
precisely the color-dipole generalization 
\cite{NZZdiffr} of the Glauber-Gribov formula \cite{Glauber,Gribov} in which no
trace of a non-Abelian intranuclear evolution with the nontrivial
attenuation eigenstate (\ref{eq:B.4}) with the eigen-cross section
(\ref{eq:B.2}) is left.

When the photon breaks into the color-singlet $q\bar{q}$
dijet, the net flow of color between the $q\bar{q}$ pair and
color-excited debris of the target nucleus is zero. Which suggests
that a rapidity gap may survive upon the hadronization, although
whether a rapidity gap in genuine inelastic events with color-singlet
$q\bar{q}$ production is stable against higher order correction or
not remains an interesting open issue. Although
the debris of the target nucleus have a zero net color charge,
the debris of color-excited  nucleons are spatially separated by a
distance of the order of the nuclear radius, which
suggests a total
excitation energy of the order of 1 GeV times $A^{1/3}$, so that
such a rapidity-gap events would look like a double diffraction with
multiple production of mesons in the nucleus fragmentation region
(for the theoretical discussion of conventional mechanisms of
diffraction
excitation of nuclei in proton-nucleus collisions see
\cite{ZollerDiffrNucl}, the experimental observation has
been reported in \cite{HELIOSNuclDiffr}).
As such, inelastic excitation of color-singlet
dijets is distinguishable from quasielastic diffractive DIS
followed by excitation and breakup of the target nucleus without
production of secondary particles.

Making use of the Sylvester expansion (\ref{eq:3.17})-(\ref{eq:3.18}),
and the
eigenstates (\ref{eq:B.3}),(\ref{eq:B.4}), one readily obtains
\bea
&& \sigma_{in}(A^*(q\bar{q})_1) = \int d^2\bb \nonumber\\
&&\times \langle \gamma^*|\left\{ (1 -\exp\left[-\sigma(\br)
T(\bb)\right]) - {N_c^2-1 \over N_c^2} (1 -\exp\left[-{1\over
2}\Sigma_2 T(\bb)\right])\right\}|\gamma^* \rangle\, ,
\label{eq:B.6} 
\eea 
\bea \sigma_{in}(A^*(q\bar{q})_8) = {N_c^2-1
\over N_c^2} \int d^2\bb \langle \gamma^*|\left\{1
-\exp\left[-{1\over 2}\Sigma_2 T(\bb)\right]\right\}|\gamma^*
\rangle \, . 
\label{eq:B.7} 
\eea 
which depend on the whole non-Abelian spectrum
of attenuation eigenstates.

Several features of the result (\ref{eq:B.6}) are noteworthy.
First, the color neutralization of the $q\bar{q}$ pair after the
first inelastic interaction requires at least one more secondary
inelastic interaction, and  an expansion of the integrand
of $\sigma_{in}(A^*(q\bar{q})_1)$
starts with the term quadratic in the optical thickness: 
\bea
\left\{(1 -\exp\left[-\sigma(\br) T(\bb)\right]) - {N_c^2-1 \over
N_c^2} (1 -\exp\left[-{1\over 2}\Sigma_2 T(\bb)\right])\right\} =
{1\over 2 (N_c^2-1)}\sigma^2(\br) T^2(\bb)+... 
\label{eq:B.8}
\eea 
Second, in the large-$N_c$ limit the color octet state tends
to oscillate in color remaining in the octet state. This is
clearly seen from (\ref{eq:B.8}). Third, in the limit of an
opaque nucleus 
\bea 
\sigma_{in}(A^*(q\bar{q})_1) = {1 \over
N_c^2}\int d^2\bb \langle \gamma^*|\left\{1
-\exp\left[-\sigma(\br) T(\bb)\right]\right\}| \gamma^* \rangle =
{1\over N_c^2} \sigma_{in} 
\label{eq:B.9} 
\eea 
and remains a
constant fraction of DIS in contrast to the quasielastic
diffractive DIS or inelastic diffractive excitation of a nucleus,
the cross sections of which vanish for an opaque nucleus
\cite{NZZdiffr,ZollerDiffrNucl}.

An analysis of the single-parton, alias the single-jet,
inclusive cross section is
quite similar. In this case we integrate over the momentum
$\bp_-$ of the antiquark jet, so that $\bb_-'=\bb_-$. The
corresponding matrix $\sigma_4$ has the eigenvalues
\bea
\Sigma_1 = \sigma(\br-\br')
\label{eq:B.10}
\eea
\bea
\Sigma_2 = {N_c^2 \over N_c^2-1} \left[\sigma(\br)+\sigma(\br')\right]
-{1\over N_c^2-1}\sigma(\br-\br')
\label{eq:B.11}
\eea
with exactly the same eigenstates $|f_1\rangle$ and $|f_2\rangle$ as
given by eqs. (\ref{eq:B.3}),(\ref{eq:B.4}). Again, the cross section
of genuine inelastic DIS corresponds to projection onto the eigenstate
$|f_1\rangle$, so that
\bea {d \sigma_{in}\over d^2\bb d^2\bp dz }   &=&  {1
\over (2\pi)^2}
 \int d^2\br' d^2\br
\exp[i\bp(\br'-\br)]\Psi^*(Q^2,z,\br')\Psi(Q^2,z,\br)\nonumber\\
&\times& \left\{\exp[-{1\over 2}\Sigma_1 T(\bb)]-
\exp[-{1\over 2}[\sigma(\br)+\sigma(\br')]T(\bb)]\right\}\nonumber\\
&=&  {1
\over (2\pi)^2}
 \int d^2\br' d^2\br
\exp[i\bp(\br'-\br)]\Psi^*(Q^2,z,\br')\Psi(Q^2,z,\br)\nonumber\\
&\times& \left\{\exp[-{1\over 2}\sigma(\br-\br') T(\bb)]-
\exp[-{1\over 2}[\sigma(\br)+\sigma(\br')]T(\bb)]\right\} \, .
\label{eq:B.12} 
\eea 
which is precisely an eq. (10) of ref.
\cite{Saturation}. 

At this point we emphasize that for the fundamental reason 
that the relevant final state is precisely the 
eigenstate $|f_1\rangle$,  
the calculation 
of the integrated inelastic cross section (\ref{eq:B.5}), as well
as the one--particle inclusive inelastic spectrum (\ref{eq:B.12}), 
are both essentially Abelian problems, and the
final result (\ref{eq:B.12}) is identical, apart from
the very different notations, to that derived 
by one of the authors (BGZ) for the propagation of 
relativistic positronium
in dense media \cite{BGZpositronium}. As one
could see from an inspection of the relevant four--parton states,
all contributions from the propagation of color--octet dipoles
cancel out, and the results could have been obtained from studying
the propagation of color--singlet dipoles without any reference to
the full cross section matrix $\sigma_4$. Our formalism makes
these cancellations nicely explicit. These 
quasi--Abelian problems have been studied also in 
\cite{Mueller,Wiedemann}.

\section*{Appendix C: Weizs\"acker-Williams glue of spatially
overlapping  nucleons}

According to \cite{Saturation,NSSdijet} the multiple convolutions
$f^{(j)}(\bkappa )$ have a meaning of the collective unintegrated
gluon SF of $j$ nucleons at the same impact
parameter the Weizs\"acker-Williams gluon fields of which overlap
spatially in the Lorentz contracted nucleus. These convolutions 
can also be viewed as a random walk in which $f(\bkappa )$
describes the single walk distribution. 

To the lowest order in
pQCD, the large $\bkappa ^2$ behavior is 
$
f(\bkappa ) \propto
\alpha_S(\bkappa ^2)/\kappa^4 $, the phenomenological study of
the differential glue of the
proton in \cite{INDiffGlue} suggests a useful  large-$\bkappa^2$
approximation $
f(\bkappa ) \propto {1/ (\bkappa^2)^{\gamma}}
$ 
with the exponent $\gamma \sim 2$ (
a closer inspection of numerical results in \cite{INDiffGlue}
gives $\gamma \approx 2.15$ at $x=10^{-2}$). The QCD
evolution effects enhance $f(\bkappa )$ at large $\bkappa ^2$, the
smaller is $x$ the stronger is an enhancement. 

Because $f(\kappa)$
decreases very slowly, we encounter a manifestly non-Gaussian
random walk. For instance, as was argued in \cite{NSSdijet}, a
$j$-fold walk to large $\bkappa^2$ is realized by one large walk,
$\bkappa_1^2 \sim \bkappa^2$, accompanied by $(j-1)$ small walks.
We simply cite here the main result \cite{NSSdijet} 
\beq
f^{(j)}(\bkappa )=j\cdot f(\bkappa )\left[1+ {4\pi^2(j-1)\gamma^2
\over N_c\sigma_0\bkappa^2} \cdot G(\bkappa^2)\right]\, ,
\label{eq:C.1} 
\eeq 
where $G(\bkappa^2)$ is the conventional
integrated gluon SF.
Then the hard tail of unintegrated nuclear
glue per bound nucleon, $f_{WW}(\bb,\bkappa )=
\phi_{WW}(\nu_A(\bb),\bkappa)/\nu_A(\bb)$, can 
be calculated parameter free: 
\bea
f_{WW}(\bb,\bkappa ) &=& {1\over \nu_A(\bb)}\cdot
\sum_{j=1}^{\infty} 
w_{j}(\nu_A(\bb))j f^{(j)}(\bkappa )\left[1+
{4\pi^2\gamma^2\over N_c\sigma_0\bkappa^2}
\cdot (j-1)G(\bkappa^2)\right] \nonumber\\
&=& f(\bkappa )\left[1+ {2 C_A\pi^2\gamma^2\alpha_S(r)T(\bb)\over
C_F N_c \bkappa^2}
G(\bkappa^2)\right]\, .
\label{eq:C.2}
\eea
In the hard regime the differential nuclear glue is not
shadowed, furthermore, because of the manifestly positive-valued
and model-independent nuclear higher twist correction it
exhibits a nuclear antishadowing property \cite{NSSdijet}.

Now we present the arguments in favor of the scaling small-$\bkappa$ behavior
\beq
f^{(j)}(\bkappa^2) \approx {1\over Q_j^2} \xi({\bkappa^2 \over Q_j^2})
\approx {1 \over \pi} {Q_j^2 \over (\bkappa^2+ Q_j^2)^2} \, 
\label{eq:C.3}
\eeq
with 
\beq
Q_j^2 \approx j Q_{0}^2\, .
\label{eq:C.4}
\eeq
In an evolution of $f^{(j)}(\bkappa )$ with $j$ at moderate $\bkappa ^2$,
\beq
f^{(j+1)}(\bkappa ^2) =\int d^2\bk f(\bk^2)f^{(j)}((\bkappa -\bk)^2)\, ,
\label{eq:C.5}
\eeq
the function $f(\bk^2)$ is a steep one compared to a smooth and
broad $f^{(j)}((\bkappa -\bk)^2)$, so that we can expand
\bea
f^{(j)}((\bkappa -\bk)^2)=f^{(j)}(\bkappa ^2) +
{df^{(j)}(\bkappa ^2) \over d\bkappa^2}[\bk^2 -2\bkappa \bk]
+{1\over 2}{d^2f^{(j)}(\bkappa^2) \over( d\bkappa^2)^2}4(\bkappa \bk)^2
\nonumber\\
\Longrightarrow f^{(j)}(\bkappa^2) +
\bk^2\left[{df^{(j)}(\bkappa ^2) \over d\bkappa^2}
+\bkappa^2 {d^2f^{(j)}(\bkappa ^2) \over( d\bkappa^2)^2}\right]
=f^{(j)}(\bkappa ^2) + \bk^2
{d \over d\bkappa^2}\left[\bkappa^2{df^{(j)}(\bkappa ^2) \over d\bkappa^2}
\right]\, .
\label{eq:C.6}
\eea
Here $\Longrightarrow$ indicates an azimuthal averaging. The expansion
(\ref{eq:C.6}) holds for for $\bk^2 \lsim Q_j^2$ and upon the $d^2\bk$
integration in (\ref{eq:C.5}) we obtain 
\bea
f^{(j+1)}(\bkappa^2)=f^{(j)}(\bkappa^2) + g(j){d \over d\bkappa^2}
\left[\bkappa^2{df^{(j)}(\bkappa ^2) \over d\bkappa^2}
\right]\, ,
\label{eq:C.7}
\eea
where
\beq
g(j) = \int^{Q_j^2}d^2\bk \bk^2f(\bk^2) =
{ 4\pi^2 \over N_c\sigma_{0}}G(Q_j^2)\, .
\label{eq:C.8}
\eeq
It is a smooth function of $j$. One can readily check
that our approximation
preserves the normalization condition $\int d^2\bkappa f^{(j)}(\bkappa)=1$.

For small $\bkappa^2$ and large $j$ the recurrence relation (\ref{eq:C.8})
amounts to the differential equation
\bea
{ Q_{j+1}^2 - Q_j^2  \over  Q_{j+1}^2 Q_j^2} = {1\over Q_j^4}\cdot
{d Q_{j}^2 \over dj} =
-{1\over Q_j^4}{\xi'(0) \over \xi(0)}g(j)
\label{eq:C.9}
\eea
with the solution
\beq
Q_{j}^2 = - {\xi'(0))\over \xi(0)}
\int^{j} dj'g(j') \approx  - jg(j) {\xi'(0)\over \xi(0)}
\label{eq:C.10}
\eeq
The expansion (\ref{eq:C.6}) holds up to the terms $\propto \bkappa^2$
and its differentiation at $\bkappa^2=0$ gives a similar constraint
on the $j$-dependence of $Q_{j}^2$.


\begin{figure}[!t]
\begin{center}
   \epsfig{file=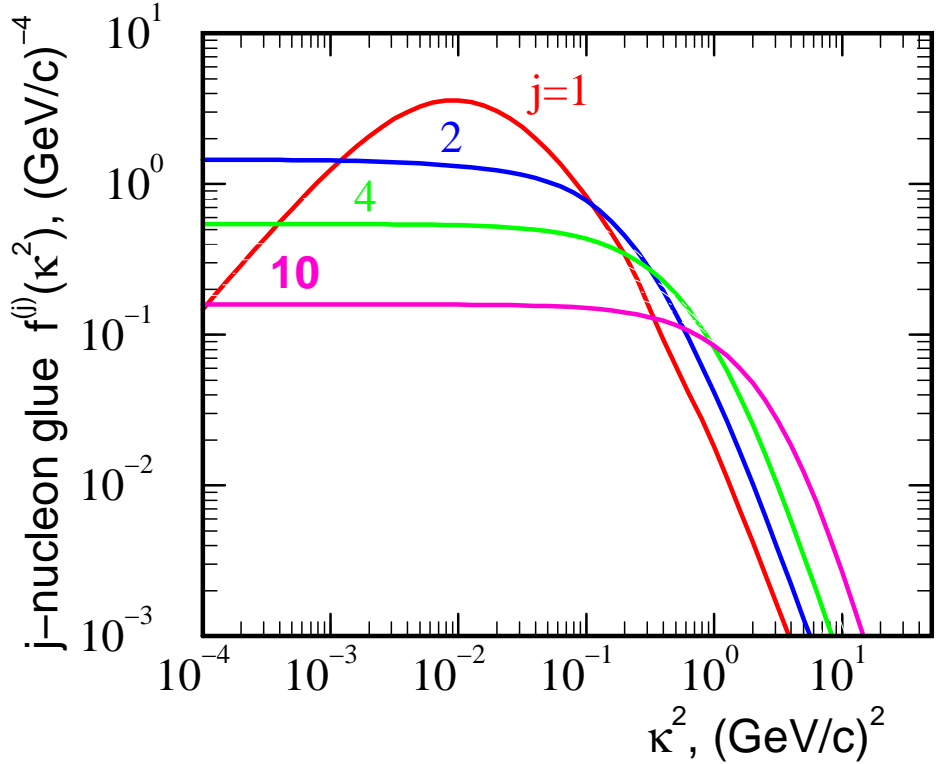,width=11cm, angle=270}
\caption{\it The nuclear dilution for soft momenta and broadening for hard
momenta of the collective glue of $j$
overlapping nucleons $f^{(j)}(\bkappa)$.  The 
numerical results are for DIS at $x=0.01$ and the input
unintegrated gluon SF of the proton is taken from ref.
\protect{\cite{INDiffGlue}}
}
\end{center}
\end{figure}

We notice that the expansion of the plateau with $j$ entails a dilution
of the differential collective glue $f^{(j)}(\bkappa^2)$ in the
plateau region, $f^{(j)}(\bkappa^2 \lsim Q_j^2) \propto 1/Q_{j}^2 \propto
1/j$.
We conclude by the observation that when extended to $\bkappa^2 \gsim
Q_{j}^2$, the parameterization (\ref{eq:C.5}), (\ref{eq:C.4}) has the
behavior $jQ_{0}^2/(\bkappa^2)^2$ which nicely matches the $j$-dependence
of the leading twist term in the hard asymptotics (\ref{eq:C.3}).

For a heavy nucleus the dominant contribution to the expansion
(\ref{eq:4.6})  comes from $j\approx \nu_A(\bb)$, so that 
\bea
\phi_{WW}(\nu_{A}(\bb),\bkappa) 
\approx {1\over \pi}{ Q_A^2(\bb) \over (\bkappa^2+ Q_A^2(\bb))^2}\, ,
\label{eq:C.11}
\eea
where (\ref{eq:C.10}) gives the width of the plateau 
\bea
Q_A^2(\bb) \approx  2\nu_{A}(\bb)g(\nu_A(\bb)) \approx
{ 4\pi^2 \over N_c} \alpha_S(Q_A^2)G(Q_A^2) T(\bb)\, .
\label{eq:C.12}
\eea
The explicit dependence on the soft parameter $\sigma_{0}$ manifest
in (\ref{eq:C.8}) cancels out in (\ref{eq:C.12}). For DIS within
the saturation domain, $Q^{2} \lsim Q_{A}^2$, the strong coupling
in (\ref{eq:4.7}) must be taken at $r \sim 1 /Q_A$, and the 
r.h.s. of (\ref{eq:C.12})  
exhibits only weak dependence on the infrared parameters through
the $Q_A^2$ dependence of the running strong coupling and 
scaling violations in the gluon
SF of the nucleon. For instance, at $x=10^{-2}$ 
the numerical results \cite{INDiffGlue}
for $G(Q^2)$ correspond to a nearly $Q^2$ independent
 $ \alpha_S(Q^2)G(Q^2)\approx 1$. For average DIS on a heavy nucleus
\bea
\langle  T(\bb) \rangle \approx {3\over 4} T(0)\approx 
{9 \over 8\pi r_0^2}A^{1/3}
\label{eq:C.13}
\eea
where $r_0 \approx 1.1$fm.  For lighter nuclei with the Gaussian 
density profile $\langle  T(\bb) \rangle \approx {1\over 2} T(0)$. 
Then for $N_c=3$ and
$A^{1/3}=6$ eqs.~(\ref{eq:C.12}) and (\ref{eq:C.13}) give 
$\langle Q_A^2(\bb)\rangle \approx 0.8$ $(GeV/c)^2$.

The utility of an approximation (\ref{eq:C.3}), (\ref{eq:C.4}) is
illustrated by fig.~7, where we show the $j$-dependence of the
collective glue of $j$ overlapping nucleons 
calculated for the unintegrated gluon SF of the proton
from ref. \cite{INDiffGlue}. For interaction of $q\bar{q}$ color 
dipoles in average DIS off the gold,
$^{197}Au$, target we find $\langle Q_{3A}^2(\bb)\rangle 
\approx 0.9$ $(GeV/c)^2$ in
good agreement with the above estimate from (\ref{eq:C.12}). For the
$q\bar{q}g$ Fock states of the photon the leading $\log Q^2$ configurations
correspond to small $q\bar{q}$ pairs which act as a color-octet
gluon \cite{NZ94} and for such gluon-gluon color dipoles
$\langle Q_{8A}^2(\bb)\rangle  \approx 2.1$ $(GeV/c)^2$. We note in passing
that the standard collinear splitting sets in, and the DGLAP
evolution \cite{Dokshitser,DGLAP} becomes applicable to nuclear structure function, 
only at $Q^2 \gg \langle Q_{8A}^2(\bb)\rangle $. 
\pagebreak\\


\begin{thebibliography}{299}

\bibitem{Textbook}
E. Leader and E. Predazzi, Introduction to Gauge Theories and
Modern Particle Physics, v.1, Cambridge University Press,
Cambridge, 1996; G. Sterman, An Introduction to Quantum Field
Theory, Cambridge University Press, Cambridge, 1993.




\bibitem{Mueller} 
A.H. Mueller, {\sl Nucl. Phys.} {\bf B558} (1999) 285; Lectures at
the Carg\`{e}se Summer School, August 6-18, 2001,
\texttt{arXiv:hep-ph/0111244}.


\bibitem{Mueller1}
A.H. Mueller, {\sl Nucl. Phys.} {\bf B335} (1990) 115.

\bibitem{McLerran}
L. McLerran and R. Venugopalan, {\sl Phys. Rev.} {\bf  D49} (1994)
2233; {\bf  D55} (1997) 5414; E. Iancu, A. Leonidov and L.
McLerran, Lectures at the Carg\`{e}se Summer School, August 6-18,
2001, \texttt{arXiv:hep-ph/0202270}.


\bibitem{Saturation}
N.N. Nikolaev, W. Sch\"afer, B.G. Zakharov, V.R. Zoller. {\sl JETP
Lett.} {\bf 76} (2002) 195.

\bibitem{NSSdijet}
N.N. Nikolaev, W. Sch\"afer and G. Schwiete, {\sl JETP Lett.} {\bf
72} (2000) 583; {\sl Pisma Zh. Eksp. Teor. Fiz.} {\bf 72} (2000)
583; {\sl Phys. Rev.} {\bf D63} (2001) 014020.

\bibitem{BFKL} 
L.N.Lipatov, {\em Sov. J. Nucl. Phys} {\bf 23} (1976) 338;
E.A.Kuraev, L.N.Lipatov and V.S.Fadin, {\em Sov. Phys. JETP} 
{\bf 44} (1976) 443;
 {\it ibid.} {\bf 45} (1977) 199; Ya.Ya.Balitsky and L.N.Lipatov,
 {\em Sov. J. Nucl. Phys} {\bf 28} (1978) 822


\bibitem{LIYaF}
  I.P. Ivanov, N.N. Nikolaev, W.
Sch\"afer, B.G. Zakharov and V.R. Zoller, Lectures
on Diffraction and Saturation of Nuclear Partons in DIS
off Heavy Nuclei, Proceedings of 36-th Annual Winter School
on Nuclear and Particle Physics and 8-th
St. Petersburg School on Theoretical Physics, St. Petersburg, Russia,
25 Feb - 3 Mar 2002.
\texttt{arXiv: hep-ph/0212161}

\bibitem{Conferences}
I.P. Ivanov, N.N. Nikolaev, W.
Sch\"afer, B.G. Zakharov and V.R. Zoller, High Density QCD, Saturation and 
Diffractive DIS. Invited talk at the 
NATO Advanced Research Workshop on Diffraction 2002, Alushta, 
Ukraine, 31 Aug - 6 September 2002.
e-Print Archive: hep-ph/0212176;
Diffractive Hard Dijets and Nuclear Parton Distributions, 
Proceedings of the Workshop on Exclusive Processes at High Momentum
     Transfer, Jefferson Lab, May 15-18, 2002. Editors A. Radyushkin 
and P. Stoler, World Scientific, 2002, pp. 205-213;
High Density QCD and Saturation of Nuclear Partons, 
Proceedings of the Conference on Quark Nuclear Physics (QNP'2002), 
June 9-14, J\"ulich,
     Germany, editors C. Elster and Th. Walcher, {\sl Eur. Phys. J}
(2003) in print, e-Print Archive: hep-ph/0209298;  
High Density QCD, Saturation and Diffractive DIS. Plenary talk
at the International Symposium on Multiparticle Dynamics (ISMD'2002),
Alushta, Ukraine, 8-14 September 2002, ed. G. Kozlov and
A. Sissakian, World Scientific (2003) in print.


\bibitem{ggFusion}
 T. Ahmed  et al. (H1 Collaboration), {\sl Nucl. Phys.} {\bf
B445}, 195 (1995) and references
therein.


\bibitem{Azimuth}
 A. Szczurek, N.N. Nikolaev, W. Sch\"afer, J. Speth,
{\sl Phys. Lett.} {\bf B500} , 254 (2001)


\bibitem{Forshaw}
J.~R.~Forshaw and R.~G.~Roberts,
Phys.\ Lett.\ B {\bf 335}, 494 (1994); A.~J.~Askew, D.~Graudenz,
J.~Kwiecinski and A.~D.~Martin,
Phys.\ Lett.\ B {\bf 338}, 92 (1994); J.~Kwiecinski, A.~D.~Martin
and A.~M.~Stasto,
Phys.\ Lett.\ B {\bf 459}, 644 (1999);

\bibitem{NZ91} 
N.N.~Nikolaev and B.G.~Zakharov, {\it Z. Phys.} {\bf C49} (1991)
607


\bibitem{NZZdiffr}
N.N. Nikolaev, B.G. Zakharov and V.R. Zoller, {\sl Z. Phys.} {\bf A351} (1995) 435.

\bibitem{BGNPZshad}
V. Barone, M. Genovese, N.N. Nikolaev,E. Predazzi and
B.G. Zakharov. {\sl Z. Phys.} {\bf C58} (1993) 541

\bibitem{NZfusion}
N.N. Nikolaev and V.I. Zakharov, {\sl Sov. J. Nucl. Phys.} {\bf
21} (1975) 227; [{\sl Yad. Fiz.} {\bf 21} (1975) 434]; {\sl Phys.
Lett.} {\bf  B55} (1975) 397.


\bibitem{NPZcharm}
N.N. Nikolaev, G. Piller and B.G. Zakharov.
{\sl J. Exp. Theor. Phys.} {\bf  81} (1995) 851;
{\sl Z. Phys.} {\bf A354} (1996) 99


\bibitem{LPM}
B.G. Zakharov, {\sl JETP Lett.} {\bf 63} (1996) 952; {\sl JETP Lett.}
{\bf 65} (1997) 615; {\sl Phys. Atom. Nucl.} {\bf 61} (1998) 838;


\bibitem{INDiffGlue}
I.P.Ivanov and N.N.Nikolaev, {\sl Phys. Atom. Nucl.} {\bf 64}, 753 (2001),
{\sl Yad. Fiz. } {\bf 64}, 813 (2001); {\sl Phys. Rev.} {\bf D65} 054004 (2002).




\bibitem{RHIC_STAR}
 C. Adler, et al. (STAR Collaboration), {\sl Phys. Rev. Lett.}
{\bf 90}, 082302 (2003)


\bibitem{Glauber}
R. J. Glauber, in {\sl Lectures in Theoretical Physics,}
 edited by W. E. Brittin et al. (Interscience Publishers, Inc., New York, 1959), Vol. 1, p. 315. 

\bibitem{Gribov}
V.N. Gribov, {\sl Sov. Phys. JETP} {\bf  29} (1969) 483; {\sl
Zh. Eksp. Teor. Fiz.} {\bf 56} (1969) 892.


\bibitem{NZ92}  
N.N. Nikolaev and B.G. Zakharov,
{\sl Z. Phys.} {\bf C53} (1992) 331.

\bibitem{NZ94} 
N.N.Nikolaev and B.G.Zakharov, {\sl  J. Exp. Theor. Phys.} {\bf
78} (1994) 806; [{\sl Zh. Eksp. Teor. Fiz.} {\bf 105} (1994)
1498]; {\sl Z. Phys.} {\bf C64} (1994) 631.

\bibitem{NZZlett}
N.N. Nikolaev, B.G. Zakharov and V.R. Zoller, {\sl JETP Lett.}
{\bf 59} (1994) 6

\bibitem{NZglue}
N.N. Nikolaev and B.G. Zakharov. {\sl Phys. Lett.} {\bf B332}
(1994) 184

\bibitem{BGNPZunit}
V. Barone, M. Genovese, N.N. Nikolaev,E. Predazzi and
B.G. Zakharov. {\sl Phys. Lett.} {\bf B326} (1994) 161


\bibitem{Andersson}
B.~Andersson {\it et al.}  (Small x Collaboration),
{\sl Eur.\ Phys.\ J. } {\bf C25}, 77 (2002)

\bibitem{GNZ95}
M. Genovese, N.N. Nikolaev and B.G. Zakharov, {\sl J. Exp. Theor.
Phys.} {\bf 81} (1995) 633; {\sl Zh. Eksp. Teor. Fiz.} {\bf  108}
(1995) 1155

\bibitem{H1gap}                             
 C. Adloff et al. (H1 Collab.), {\sl  Z. Phys.} {\bf  C76} (1997) 613 

\bibitem{ZEUSgap}
 J.Breitweg et al. (ZEUS Collaboration),
{\sl Europ. Phys. J.} {\bf  C6} (1999) 43.
     


\bibitem{BGZpositronium}
B.G. Zakharov, {\sl Sov. J. Nucl. Phys.} {\bf 46} (1987) 92; {\sl Yad. Fiz.} 
{\bf 46} (1987) 148.


\bibitem{NSZdist}
N.N. Nikolaev, J. Speth, B.G. Zakharov,
{\sl J. Exp. Theor. Phys.} {\bf 82} (1996) 1046; {\sl
Zh. Eksp. Teor. Fiz.} {\bf 109} (1996) 1948


\bibitem{Dokshitser}
Yu.L.~Dokshitser, {\it Sov. Phys. JETP} {\bf 46} (1977) 641;
{\sl Zh. Eksp. Teor. Fiz.} {\bf 73}, 1216 (1977);
Yu.L. Dokshitzer, D. Diakonov, S.I. Troian, {\sl Phys. Rept.} {\bf 58}, 269 (1980)
 
\bibitem{LUND}
T.Sj\"ostrand at al., {\sl Comp. Phys. Commun.} {\bf 135} (2001) 238

\bibitem{Pions}
 N.N. Nikolaev, W.
Sch\"afer, B.G. Zakharov and V.R. Zoller, paper in preparation.

\bibitem{ZollerDiffrNucl}
V.R. Zoller, {\sl Z. Phys.} {\bf C51} (1991) 659; M.A. Faessler,
{\sl Z. Phys.} {\bf C58} (1993) 567.

\bibitem{HELIOSNuclDiffr}
T.Akesson et al. (HELIOS Collab.), {\sl Z. Phys.} {\bf C49} (1991) 355

\bibitem{DeJaeger}
H. De Vries, C.W. De Jaeger, C. De Vries, {\sl Atomic Data and 
Nuclear Data Tables} {\bf 36}, 495 (1987)

\bibitem{Wiedemann}
U.~A.~Wiedemann,
{\sl Nucl.\ Phys.\ }  {\bf B582}, 409 (2000) 

\bibitem{DGLAP}
V.N.~Gribov and L.N.~Lipatov, {\it Sov. J. Nucl. Phys.} {\bf 15} (1972)
438; L.N.~Lipatov, {\it Sov. J. Nucl. Phys.} {\bf 20} (1974) 181;
G.~Altarelli and G.~Parisi, {\it Nucl. Phys.} {\bf B126} (1977) 298.

\end{thebibliography}
\end{document}